\documentclass[12pt]{article}

\usepackage{amsmath,amssymb,graphicx} %use drftcite in draftmode
\usepackage{epsf}
\usepackage{pstricks}
\usepackage{cite}
\newcommand{\beq}{\begin{eqnarray}}% can be used as {equation} or  {eqnarray}
\newcommand{\eeq}{\end{eqnarray}}

\newcommand{\centeron}[2]{{\setbox0=\hbox{#1}\setbox1=\hbox{#2}\ifdim

\wd1>\wd0\kern.5\wd1\kern-.5\wd0\fi \copy0

\kern-.5\wd0\kern-.5\wd1\copy1\ifdim\wd0>\wd1
                                      \kern.5\wd0\kern-.5\wd1\fi}}
\newcommand{\ltap}{\>\centeron{\raise.35ex\hbox{$<$}}
                              {\lower.65ex\hbox{$\sim$}}\>}
\newcommand{\gtap}{\>\centeron{\raise.35ex\hbox{$>$}}
                              {\lower.65ex\hbox{$\sim$}}\>}

\newcommand\ZZ{\hbox{\zfont Z\kern-.4emZ}}
\font\zfont = cmss10 %scaled \magstep1

\def\beqa{\begin{eqnarray}}
\def\eeqa{\end{eqnarray}}

\def\ben{\begin{enumerate}}
\def\een{\end{enumerate}}
\def\bei{\begin{itemize}}
\def\eei{\end{itemize}}

\begin{document}
\begin{titlepage}
%\begin{flushright}
%{\tt hep-ph/yymmnn}
%\end{flushright}

\vskip1.5cm
\begin{center}
{\huge \bf A Model of Lepton Masses}\\
\vspace{.15cm}
{\huge \bf from a Warped Extra Dimension} \vspace*{0.1cm}
\end{center}
\vskip0.2cm

\begin{center}
{\bf Csaba Cs\'aki$^{a,b}$, C\'edric Delaunay$^{c,d}$, Christophe
Grojean$^{c,d}$ \\ and Yuval Grossman$^a$}

\end{center}
\vskip 8pt

\begin{center}
$^{a}$ {\it Institute for High Energy Phenomenology\\
Newman Laboratory of Elementary Particle Physics\\
Cornell University, Ithaca, NY 14853, USA } \\
\vspace*{0.3cm}
$^b$ {\it  Kavli Institute for Theoretical Physics\\
University of California, Santa Barbara, CA 93106, USA} \\
\vspace*{0.3cm}
$^c$ { \it CERN Theory Division, CH-1211 Geneva 23, Switzerland} \\
\vspace*{0.3cm}
$^d$ {\it Institut de Physique Th\'eorique,
CEA, IPhT, F-91191 Gif-sur-Yvette, France}
\vspace*{0.3cm}

{\tt  csaki@lepp.cornell.edu,
cedric.delaunay@cea.fr, \\ christophe.grojean@cern.ch, yuvalg@lepp.cornell.edu}
\end{center}

\vglue 0.3truecm

\begin{abstract}
\vskip 3pt \noindent

In order to explain the non-hierarchical neutrino mixing angles and
the absence of lepton flavor violating processes in the context of
warped extra dimensions one needs to introduce bulk flavor
symmetries. We present a simple model of lepton masses and mixings in
RS models based on the A$_4$ non-abelian discrete symmetry. The
virtues of this choice are: $(i)$ the natural appearance of the
tri-bimaximal mixing pattern; $(ii)$ the complete absence of
tree-level flavor violations in the neutral sector; $(iii)$ the
absence of flavor gauge bosons; $(iv)$ the hierarchies in the charged
lepton masses are explained via wave-function overlaps. We present the
minimal field content and symmetry breaking pattern necessary to
obtain a successful model of this type. The bounds from electroweak
precision measurements allow the KK mass scale to be as low as $\sim$ 3~TeV. 
Tree-level lepton flavor violation is absent in this model, while the loop induced
$\mu\to e\gamma$ branching fraction is safely below the experimental bound.

\end{abstract}

\end{titlepage}

%\newpage

%\renewcommand{\thefootnote}{(\arabic{footnote})}

%%%%%%%%%%%%%%%%%%%%%%%%%%%%%%%%%%%%%%%%%%%%%%%%%%%%%%
%%%%%%%%%%%%%%%%%%%%%%%%%%%%%%%%%%%%%%%%%%%%%%%%%%%%%%
\section{Introduction}
\label{sec:Intro} \setcounter{equation}{0} \setcounter{footnote}{0}
%%%%%%%%%%%%%%%%%%%%%%%%%%%%%%%%%%%%%%%%%%%%%%%%%%%%%%
%%%%%%%%%%%%%%%%%%%%%%%%%%%%%%%%%%%%%%%%%%%%%%%%%%%%%%

Warped extra dimensions~\cite{RS} provide a simple framework for
fermion masses: exponential hierarchies are naturally generated due to
the overlap of fermion and Higgs wave functions~\cite{GN,GP},
implementing the split fermion idea of~\cite{AS}. In the simplest 5D
``anarchic" approach~\cite{HS,APS}, where both the 5D bulk masses and
the brane Yukawa couplings are assumed to be random, one generates a
hierarchy both in the 4D standard model (SM) fermion masses and their
mixing angles. This seems to fit the observed pattern of quark mixings
and masses very well, since both the CKM matrix and quark masses show
a hierarchical pattern. The lepton sector, however, is different: two
of the observed neutrino mixing angles are close to maximal, rather
than being hierarchical~\cite{pdg}. This suggests that one needs to
radically change the implementation of fermion masses in warped extra
dimensional models in order to achieve the correct lepton mixing
pattern. Instead of a fully 5D anarchical approach, it calls for partial
flavor symmetries, which will make sure that two of the neutrino mixing
angles are not small, but close to maximal. The usual hierarchies can
still be used to generate the hierarchies in charged lepton
sector. 

In fact, the necessary appearance of flavor symmetries is welcome for
these models. Anarchic 5D flavor structure necessarily gives rise to
flavor changing neutral currents (FCNC's) in the quark sector, and to
lepton flavor violation (LFV) in the lepton sector at
tree-level~\cite{GP,HS,Gustavo,APS}. While these flavor violations are
suppressed by the so-called RS-GIM mechanism~\cite{Gustavo,APS}, the
KK mass scale still has to be as large as 20 TeV in order to safely
suppress FCNC's in the quark sector~\cite{CFW}. In the lepton sector
the anarchic approach~\cite{HSleptons} imposes a bound of order 10 TeV
on the KK mass scale~\cite{AP}. These bounds imply that the theory is
out of reach for the LHC and unlikely to be useful for solving the
(little) hierarchy problem.  Reduction of these bounds require
additional flavor symmetries~\cite{XDGim,inprep}. For the leptons, we
have seen that such flavor symmetries are necessary to begin with, to
protect the mixing angles from becoming hierarchical. The aim of this
paper is to show that we can indeed use these flavor symmetries in the
lepton sector to eliminate the LFV bounds on the model, and at the
same time get a realistic pattern of masses and mixings. Recently,
an alternative
approach to lepton masses in RS was proposed in~\cite{MuChun}
(following the suggestion of~\cite{FPR}), which however does not
readily explain the appearance of the large (non-hierarchical)
neutrino mixing angles.

For our model, we pick the most popular global symmetry used in
neutrino model building, an A$_4$ discrete symmetry~\cite{Ma,AF}.
This, by the virtue of being discrete, has the added benefit that no
additional gauge bosons have to be introduced, even if the symmetry is
gauged. From the many studies of the implications of the A$_4$
symmetry in 4D models we expect that this symmetry can give the
correct mixing structure. The mass hierarchies in the charged sector
can still be generated as usual in RS models via fermion overlaps. 
The mass hierarchy in the neutrino sector is not that large, and can be readily incorporated by choosing 
$\mathcal{O}(1)$ factors in the neutrino mass matrix.
Finally, since A$_4$ is a non-abelian discrete symmetry it has the
potential of greatly reducing bounds from LFV.  The reason for that is
that by using a non-singlet representation under A$_4$ we can force
the bulk wave functions of some of the left handed fields to be
universal.

Indeed, we find that with appropriate choice of representations, the
tri-bimaximal mixing pattern~\cite{HPS} characteristic of the A$_4$
symmetry can be reproduced by the leading order operators. Higher
order terms can result in non-zero $\theta_{13}$, while maintaining
the predictions for $\theta_{12},\theta_{23}$ within the
experimentally allowed range. Most importantly, our A$_4$ based model
eliminates all tree-level sources for LFV. In fact, LFV is completely
absent in the neutral current interactions, and shows up only through
charged currents involving neutrino mixing. In the original RS
neutrino mass model of~\cite{GN} these loops turn out to be
problematic~\cite{Kitano}, since the interaction of the heavy KK
neutrinos with the SM fields is unsuppressed. In our model, by putting
the charged leptons in the bulk (and peaked on the UV brane) these
couplings are strongly suppressed and the bounds from loop induced
decays are significantly reduced. The most important experimental
bounds in the lepton sector are those from the electroweak precision
(EWP) constraints. They are, however, quite mild, with KK masses of
order $3$ TeV generically allowed as in most other RS models.

The above arguments show that introducing the discrete non-abelian
lepton flavor symmetry greatly improves over the generic RS lepton
flavor models. Moreover, they are also improving the straight 4D
implementations of A$_4$ neutrino models in several aspects.  First,
they explain the hierarchy in the charged lepton sector. Second, by
putting one A$_4$ breaking VEV on the UV brane (breaking the group to
Z$_2$), and the other on the IR (breaking it to Z$_3$) questions
regarding vacuum alignment are eliminated. Finally, the appearance of
the correct right handed neutrino mass scale (somewhat below $M_{Pl}$ and
$M_{GUT}$) can be explained by the partial compositeness of the right
handed neutrino.

The paper is organized as follows: in Section 2 we give the general
setup, introduce the A$_4$ representations and calculate the mixing
matrices at leading order. In Section 3 we show the effects of higher
dimensional operators on the mixing angles. In Section 4 we present a
numerical scan of the parameter space and discuss the
electroweak precision bounds. In Section 5 we show that LFV is
completely absent at the tree-level in this model, and estimate the
loop induced $\mu\to e\gamma$ branching fraction. We conclude in
Section 6.

%%%%%%%%%%%%%%%%%%%%%%%%%%%%%%%%%%%%%%%%%%%%%%%%%%%%%%
%%%%%%%%%%%%%%%%%%%%%%%%%%%%%%%%%%%%%%%%%%%%%%%%%%%%%%
\section{The Setup}
\label{sec:setup} \setcounter{equation}{0} \setcounter{footnote}{0}
%%%%%%%%%%%%%%%%%%%%%%%%%%%%%%%%%%%%%%%%%%%%%%%%%%%%%%
%%%%%%%%%%%%%%%%%%%%%%%%%%%%%%%%%%%%%%%%%%%%%%%%%%%%%%

We are assuming an AdS$_5$ bulk metric
\begin{equation}  \label{eq:metric}
ds^2=\left(\frac{R}{z}\right)^2 (dx_\mu dx_\nu \eta^{\mu\nu} -dz^2),
\end{equation}
with a UV~brane at $z=R$ ($R$ is also the AdS curvature scale) and
an IR~brane at $z=R'$. The magnitude of the scales is given by
$R^{-1}\sim M_{Pl}$ and $R'^{-1} \sim 1$ TeV. The electroweak gauge
group is extended to an SU(2)$_L\times$SU(2)$_R\times$U(1)$_{B-L}$
gauge symmetry in the bulk to incorporate custodial
symmetry~\cite{ADMS}. This symmetry is reduced on the UV~brane to
the SM group SU(2)$_L\times$U(1)$_Y$, while it breaks down to
SU(2)$_D\times$U(1)$_{B-L}$ on the IR~brane.

\begin{table}[t]
\begin{displaymath}
\begin{array}{|c|c|c|c|c|c|} \hline
            & SU(2)_L   & SU(2)_R & U(1)_{B-L} & A4 & Z_2\\ \hline
\Psi_L &  \Box     &  1      &    -1       & 3  & -\\ \hline
  \Psi_{e,\mu,\tau} &  1        & \Box    &  -1   & 1,1',1'' & + \\ \hline
  \Psi_{\nu}  &  1        & \Box    &    -1       & 3   & - \\ \hline
    H \ (IR)   &  \Box     & \Box    &     0      & 1  & + \\ \hline
   \phi'\ (IR)  &  1        &  1      &     0      & 3  & - \\ \hline
   \phi\ (UV) &  1        &   1     &     0      & 3  & + \\ \hline
\end{array}
\end{displaymath}
\caption{Fields and their gauge and flavor charges.}
\label{tab:fieldcontent}
\end{table}

The matter content is summarized in Table~\ref{tab:fieldcontent}. We
assume that there is a separate doublet for every SM lepton, including
the right handed neutrino: an SU(2)$_L$ doublet, $\Psi_L$, for every
left handed doublet $L$, a separate SU(2)$_R$ doublet $\Psi_{e,\mu
,\tau}$ for every right handed charged lepton, $e,\mu ,\tau$, and a
right handed doublet $\Psi_\nu$ for every right handed neutrino
$\nu$. A 5D fermion correspond to 2 Weyl fermions of opposite
chirality in 4D:
\begin{equation}
\Psi = \left( \begin{array}{c} \chi \\ \bar{\psi} \end{array}�\right).
\end{equation}
In the KK decompostion, 4D chirality follows from the boundary
conditions at the two end points of the extra dimension. The Lorentz
structure forces $\chi$ and $\psi$ to have opposite boundary
conditions while a massless mode appears only for Neumann boundary
condition at both ends. For a complete description of fermionic
boundary conditions see~\cite{CGHST}. The profile of the would-be zero
mode is then entirely dictated by the 5D mass of $\Psi$.
Conventionally, this mass is normalized as $c/R$ and for $c>1/2$
(resp. $c<-1/2$), a $\chi$-zero mode (resp. $\psi$-zero mode) is
exponentially localized on the UV~brane. In our setup, we assume that
the boundary conditions (in the absence of the localized mass terms)
are chosen as:
\begin{equation}
\begin{array}{ccc}
\Psi_L=\left(\begin{array}{cc} L & [+,+]\end{array}\right) & \Psi_{e,\mu,\tau}=\left(\begin{array}{cc} \tilde{\nu}_{e,\mu,\tau} & [+,-]\\
e,\mu,\tau & [-,-]\end{array}\right) & \Psi_\nu=\left(\begin{array}{cc} \nu & [-,-]\\
\tilde{l} & [+,-]\end{array}\right)
\end{array}
\end{equation}
where $[\pm]$ refers to a Neumann (Dirichlet) boundary condition on
both branes for the $\chi$ component, while the $\bar{\psi}$ ones
simply have the opposite conditions. Hence there is a left handed zero
mode for the left handed doublets in $\Psi_L$, and a single right
handed zero mode in $\Psi_{e,\mu ,\tau}$ and $\Psi_\nu$
each.\footnote{It is well known that the $[+-]$ boundary conditions
(for a $\chi$-type Weyl fermion) can lead to an extremely light
KK-state for $c>1/2$. Here we are safe from this problem as the right
handed zero modes which satisfy this type of boundary conditions are
localized close to the UV~brane, i.e. $c<-1/2$.}  These fields have
bulk masses (in units of the AdS curvature) given by $c_L, c_e$ and
$c_\nu$. For the general case these would be hermitian $3 \times 3$
flavor matrices. This is not the case here as we impose an A$_4\times$Z$_2$
global symmetry in the bulk of the theory.
%, with specific symmetry
%breaking VEVs on the UV~and the IR~branes. 
In order to get the correct neutrino mass spectrum, we assign the
three charged lepton doublets as well as the three right handed
neutrinos to the 3 dimensional representation of A$_4$.  Thus for
these fields there is just one common $c$-parameter each: $c_L$
and $c_\nu$. On the other hand, we assign the right handed
charged leptons to the three inequivalent one dimensional
representations of A$_4$: 1+1'+1''. Thus there are three separate
$c$'s in this sector: $c_e$, $c_\mu$ and $c_\tau$.
The purpose of the Z$_2$ symmetry is to eliminate certain
brane localized operators that would otherwise contribute to the mass
matrices at leading order.

The symmetry breaking is achieved via brane localized scalars.
Besides the SM Higgs, $H$, that is localized on the IR~brane, we assume the
following scalars to break the discrete symmetries: $\phi$ on the UV
brane and $\phi'$ on the IR~brane, both of which are in the 3 of
A$_4$.  We assume that these two
scalars develop VEVs in different directions: $\phi$ breaks A$_4$
to Z$_2$, while $\phi'$ to Z$_3$. This is achieved by the VEVs %%
\begin{equation}
\langle\phi'\rangle = (v',v',v'),\qquad 
\langle\phi\rangle = (v,0,0),\label{eqn:UVvev}
\end{equation}
in the basis where the generator corresponding to  generator $S$ of
A$_4$ is diagonal (see Appendix~A for summary on
A$_4$). Note, that once such a basis is chosen, these are
the most general VEVs which preserve Z$_3$ and Z$_2$ subgroups of
A$_4$ respectively, up to a trivial permutation of the basis.

We now write the most general Yukawa terms respecting both gauge and
flavor symmetries (in addition one needs to write localized kinetic
and potential terms for the localized scalars):
{\setlength\arraycolsep{2pt}
\begin{eqnarray}
{\cal L}_{UV}&=&-\frac{M}{2\Lambda}\psi_{\nu}\psi_{\nu}-x_\nu\frac{\phi}{2\Lambda}\psi_{\nu}\psi_{\nu}+{\rm
h.c.}+\cdots,  \\
{\cal L}_{IR}&=& -\frac{y_\nu}{\Lambda'}\overline{\Psi}_L H\Psi_\nu
-\frac{y_e}{\Lambda'^2}
\left(\overline{\Psi}_L\phi'\right)H\Psi_e-\frac{y_\mu}{\Lambda'^2}
\left(\overline{\Psi}_L\phi'\right)'H\Psi_\mu
-\frac{y_\tau}{\Lambda'^2}\left(\overline{\Psi}_L\phi'\right)''H\Psi_\tau+{\rm
h.c.}+\cdots,\nonumber 
\label{eq:boundaryaction}
\end{eqnarray}
}where $\Psi^T=\left(\chi,\bar{\psi}\right)$ and $\cdots$ stands for
higher dimensional operators. We use the notation of \cite{AF} for
writing A$_4$-invariants: $()$,$()'$ and $()''$ denote the terms of
$3\times 3$ that transform as $1$,$1'$ and $1''$ respectively,
see appendix \ref{sec:A4}.

Once the electroweak and A$_4$ symmetries are spontaneously broken,
the boundary terms lead to boundary conditions mixing all
fermions. Then, the spectrum is obtained by solving the bulk equation
of motion in the presence of these mixed boundary conditions. The
light modes, however, are quite insensitive to the boundary terms, and
so they can be treated as a small perturbation. Hence to leading order
the low energy spectrum may be obtained by using the zero mode wave
functions. This defines the Zero Mode Approximation (ZMA) whose
accuracy depends on how light the light masses are. As the largest
mass is that of the $\tau$, about 1~GeV, the ZMA turns out to be as
accurate as $m_\tau R'\simeq 10^{-3}$ for zero mode masses.

To follow the conventional RS literature we introduce the RS flavor
functions $f$ and $F$; these give the wave functions of the zero
mode fermions on the IR~and UV~branes:
\begin{equation}
\chi_{c} (z) =\frac{1}{\sqrt{R'}} \left(\frac{z}{R}\right)^2 \left(\frac{z}{R'}\right)^{-c} f_c
=\frac{1}{\sqrt{R}}  \left(\frac{z}{R}\right)^{2-c} F_c~
\ \ \textrm{and }\  ~ \psi_c (z) = \chi_{-c} (z),
\end{equation}
with
\begin{eqnarray}
f_c =\frac{\sqrt{1-2c}}{\sqrt{1-(R/R')^{1-2c}}},\qquad
F_c=\frac{\sqrt{2c-1}}{\sqrt{1-(R/R')^{2c-1}}}.
\end{eqnarray}
The IR boundary terms of~(\ref{eq:boundaryaction}) lead to the
following Dirac mass matrices for the zero mode charged leptons and
neutrinos:
\begin{equation}
\mathbf{\mathcal{M}_{D}^e} = f_{L}
\frac{v_H v'}{\sqrt{2} R' \Lambda'^2}
\left(\begin{matrix}
y_e f_{-e} & y_\mu f_{-\mu}& y_\tau f_{-\tau}\\\
y_e f_{-e} & \omega y_\mu f_{-\mu}& \omega^2 y_\tau f_{-\tau}\\
y_e f_{-e} & \omega^2 y_\mu f_{-\mu}& \omega y_\tau f_{-\tau}
\end{matrix}\right),
\label{eq:mDe}
\end{equation}
\begin{equation}\label{eqn:mDnu}
\mathbf{\mathcal{M}_{D}^\nu} =  y_\nu f_{L} f_{-\nu}
\frac{v_H}{\sqrt{2} R' \Lambda'}
\left(\begin{matrix}
1&0&0\\
0&1&0\\
0&0&1
\end{matrix}\right),
\end{equation}
where we have introduced the shorthanded notation $f_i=f_{c_i}$
and $f_{-i}=f_{-c_i}$), $\omega$ is the cubic root of unity, $\omega =e^{2\pi
i/3}$, and $v_H\sim$ 250 GeV.
The UV~terms of (\ref{eq:boundaryaction}) generate a Majorana
mass matrix for right handed neutrinos of the form: %%
\begin{equation}\label{eqn:MR}
\mathbf{\mathcal{M}_M^\nu} =  F^2_{-\nu} R^{-1}
\left(\begin{matrix}
\epsilon_s & 0 & 0 \\
0 & \epsilon_s & \epsilon_t\\
0 & \epsilon_t & \epsilon_s
\end{matrix}\right).
\end{equation}
with  $\epsilon_s\equiv M/\Lambda$ and
$\epsilon_t\equiv x_\nu v/\Lambda$. 

Note, that while the mass hierarchy in the charged lepton sector is
generated via the wave function overlaps in the usual way, the
non-degeneracy in the neutrino sector is actually due to the two
different kind of Majorana term allowed on the UV~brane. As discuss
below, the required neutrino mass hierarchy will actually require
$\epsilon_s \sim \epsilon_t$. This is more naturally achieved if the
singlet Majorana mass is actually also originating from an operator
with a singlet scalar field VEV. This could for example be enforced by
imposing an additional Z$_3$ global symmetry, and an additional scalar
field $\xi$ with VEV $\langle \xi \rangle =M$.

After integrating out the heavy right handed neutrinos, one ends up
with a seesaw type Majorana mass matrix of the left handed
neutrinos %%
\begin{eqnarray}
\mathbf{\tilde{\mathcal{M}}_{M}^\nu}&\equiv&-
\mathbf{\mathcal{M}_{D}^\nu}\cdot\left(\mathbf{\mathcal{M}_{M}^\nu}\right)^{-1}
\cdot\left(\mathbf{\mathcal{M}_{D}^\nu}\right)^T\nonumber\\
&=&-y_\nu^2\frac{v_H^2 R}{2\Lambda'^2 R'^2}
\frac{f_L^2 f_{-\nu}^2}{F_{-\nu}^2}
\left(\begin{matrix}
1/\epsilon_s & 0 & 0 \\
0 & \epsilon_s/\Delta & -\epsilon_t/\Delta\\
0 & -\epsilon_t/\Delta & \epsilon_s/\Delta
\end{matrix}\right),
\end{eqnarray}
with $\Delta\equiv \epsilon_s^2-\epsilon_t^2$. 
The diagonalization procedure is the same as that of usual A$_4$
four-dimensional models. The charged lepton mass matrix becomes diagonal
once the left-handed fields have been rotated as $L\rightarrow
\mathbf{V}\cdot L$ with %%
\begin{equation}\label{eqn:V}
\mathbf{V}=\frac{1}{\sqrt{3}}\left(\begin{matrix}
            1 & 1 & 1 \\
            1 & \omega^2 & \omega \\
            1 & \omega & \omega^2 \end{matrix}\right),
\end{equation}
a symmetric unitary matrix. We emphasize already here the most
important properties of the A$_4$ mass matrices: the left handed
rotation needed for diagonalizing the mass matrix is independent of
the actual magnitudes of the masses, and the right handed rotation is
the unit matrix (i.e. no right handed rotation is
necessary). Therefore, the charged lepton masses are %%
\begin{equation}
\mathbf{V}^*\cdot \mathbf{\mathcal{M}_{D}^e}=f_L
\frac{v_H v'}{\sqrt{2} R' \Lambda'^2}
\left(\begin{matrix}
y_e f_{-e}&0&0\\
0&y_\mu f_{-\mu}&0\\
0&0&y_\tau f_{-\tau}
\end{matrix}\right).
\label{eq:mDediag}
\end{equation}
The charged lepton mass hierarchies are due to the hierarchies on
$f_{-e,-\mu ,-\tau}$ and the A$_4$ embedding of the right handed
charged leptons allows for three different $c$'s which can be set to
generate the physical charged lepton masses. 

Next we move to the light neutrino
sector. We work in the basis of diagonal charged leptons
which is obtained by performing the rotation on the entire left-handed
doublet with $\mathbf{V}$. Then, 
the light neutrino 
Majorana mass matrix is diagonalized by the
Harrison--Perkins--Scott (HPS)
matrix~\cite{HPS} %%
\begin{equation}
U_{HPS}=\left(\begin{matrix}
\sqrt{2/3} & 1/\sqrt{3} & 0 \\
-1/\sqrt{6} & 1/\sqrt{3} & -1/\sqrt{2} \\
-1/\sqrt{6} & 1/\sqrt{3} & 1/\sqrt{2}\\
\end{matrix}\right),
\end{equation}
corresponding to $\theta_{13}=0$, $\sin^2 (2 \theta_{12})=8/9$ and
$\theta_{23}=\pi/4$.  This tri-bimaximal mixing matrix is very close
to the the best fit obtained from present oscillation data. The
neutrino mass eigenstates are
\begin{equation}
U_{HPS}^T\cdot \mathbf{V}^*\cdot \mathbf{\tilde{\mathcal{M}}_{M}^\nu}\cdot \mathbf{V}^*\cdot U_{HPS}
=-\tilde{m}
\left(\begin{matrix}
\frac{1}{\epsilon_s+\epsilon_t}&0&0\\
0&\frac{1}{\epsilon_s}&0\\
0&0&\frac{1}{\epsilon_t-\epsilon_s}
\end{matrix}\right),
\end{equation}
where the overall mass scale is set by the combination
\begin{equation}
\tilde{m}\equiv y_\nu^2\frac{v_H^2 R}{2\Lambda'^2 R'^2}
\frac{f_L^2 f_{-\nu}^2}{F_{-\nu}^2},
\end{equation}
The neutrino mass-squared splittings are given by
\begin{eqnarray}
\Delta m_{12}^2&\equiv&|m_1|^2-|m_2|^2=\left|\frac{\tilde{m}}{\epsilon_s}\right|^2\left[\frac{1}{\left(1+r\right)^2}-1\right],\\
\Delta m_{23}^2&\equiv&|m_2|^2-|m_3|^2=\left|\frac{\tilde{m}}{\epsilon_s}\right|^2\left[1-\frac{1}{\left(1-r\right)^2}\right],
\end{eqnarray}
with $r\equiv\epsilon_t/\epsilon_s$, $|\Delta m_{12}^2|=\Delta
m^2_{sol}$ and $|\Delta m_{23}^2|=\Delta m^2_{atm}$.  Combining the
last two relations we see that $r$ solves the following algebraic
equation: %%
\begin{equation}
r^3-3r-2\left(\frac{x-1}{x+1}\right)=0,
\end{equation}
where $x=\Delta m^2_{sol}/\Delta m^2_{atm}$ for $|r|<2$ or $x=-\Delta
m^2_{sol}/\Delta m^2_{atm}$ when $|r|>2$. Finally, $\epsilon_s$ is
found by inverting one of the two relations, for instance,
\begin{equation}
\epsilon_s=\frac{\tilde{m}}{\sqrt{\Delta m^2_{atm}}}
\times \left(\left|1-\frac{1}{(1-r)^2}\right|\right)^{-1/2}.
\end{equation}

When we impose the measured values~\cite{pdg} for the mass splittings
\beq \label{nudata}
\Delta m^2_{sol}\simeq 7.9\times 10^{-5}~{\rm eV}^2, \qquad\Delta
m^2_{atm}\simeq 2.6\times 10^{-3}~{\rm eV}^2,
\eeq
we find four solutions for the neutrino mass spectrum corresponding to
\beq \label{rsol}
r\approx\{0.79,1.19,-2.01,-1.99\}.
\eeq
The first (last) 2 solutions lead to a normal (inverted) hierarchy
spectrum. Another general feature of A$_4$ is a prediction of the
absolute neutrino mass scale, $\tilde m$.
%Indeed, as soon as the two splittings
%are reproduced and since the spectrum is set in terms of only two
%Majorana masses on the UV~brane, 
We find that the mass of the heaviest neutrino ends up being slightly
above the atmospheric splitting.

We close this section by presenting a set of numerical values for the
Lagrangian parameters which reproduce the mass spectra. The brane
positions are $R^{-1}=10^{19}$~GeV and $R'^{-1}=1.5$~TeV, in order to
keep the KK gauge bosons (with $m_{KK}=3-4$~TeV) in the reach of the
LHC. The Higgs VEV turns out to be $v_H=255.5$~GeV, which is obtained
after matching the bulk gauge couplings such that the weak boson
masses as well as the fine-structure constant at the $Z$ pole take
their physical values: $m_W=80.40$~GeV, $m_Z=91.19$~GeV and
$\alpha_{em}^{-1}(m_Z)=128$. As usual~\cite{Cacciapaglia:2006mz}, we get a Higgs VEV
slightly above the SM value due to the suppression of the $W,Z$ wave
functions on the IR brane and the additional contributions to the
gauge boson masses from the wave function curvature terms.

The charged lepton masses are reproduced using the following choice of
parameters: $c_L=0.51,\ c_e=-0.75,\ c_\mu=-0.59,\ c_\tau=-0.51$,
$y_e=1.53,\ y_\mu=1.55,\ y_\tau=3.04$. Indeed, together with
$\Lambda'=R'^{-1}$ and $v'R'=0.1$, one gets $m_e=0.511$ MeV,
$m_\mu=106$ MeV and $m_\tau=1.77$ GeV. We show later on that the model
defined with this set of parameters passes all leptonic electroweak
precision bounds while perturbation theory remains under control up to
$E=3 m_{KK}$.  As we saw above, in the neutrino sector the solar and
atmospheric mass splittings, Eq.~(\ref{nudata}), reduce the ratio $r$
to only four possible values, Eq.~(\ref{rsol}). The corresponding mass
spectra, as well as the Majorana masses on the UV brane (for
$\Lambda=R^{-1}$, $x_\nu=1$ and $c_\nu=-0.37$), are reported in Table
\ref{tab:egmasses}.

\begin{table}[t]
\begin{displaymath}
\begin{array}{|c|c|c|c|c|c|} \hline
      r      & m_1   & m_2  & m_3  & MR & vR \\ \hline
-2.01 &  53     &  54    &  18  & -0.015 & 0.030\\ \hline
-1.99 &  55       & 54    &   18  & -0.015 & 0.029 \\ \hline
 0.79 &  6.0       & 11    &    52   & 0.074  & 0.059 \\ \hline
 1.19 &  4.5     & 10    &    52    &  0.079 & 0.095 \\ \hline
\end{array}
\end{displaymath}
\caption{Approximate numerical values of the neutrino masses and UV
VEVs for the 4 possible solutions of
$r\equiv\epsilon_t/\epsilon_s$. The masses are given in units of
$10^{-3}$ eV.} \label{tab:egmasses}
\end{table}

%%%%%%%%%%%%%%%%%%%%%%%%%%%%%%%%%%%%%%%%%%%%%%%%%%%%%%
%%%%%%%%%%%%%%%%%%%%%%%%%%%%%%%%%%%%%%%%%%%%%%%%%%%%%%
\section{Higher Order Corrections and Perturbativity Bounds}
\label{sec:HOcorr} \setcounter{equation}{0} \setcounter{footnote}{0}
%%%%%%%%%%%%%%%%%%%%%%%%%%%%%%%%%%%%%%%%%%%%%%%%%%%%%%
%%%%%%%%%%%%%%%%%%%%%%%%%%%%%%%%%%%%%%%%%%%%%%%%%%%%%%

After spontaneous breaking of A$_4$ the lowest dimensional boundary
terms of (\ref{eq:boundaryaction}) generate the tri-bimaximal pattern
for the mixing angles as well as the charged lepton and neutrino mass
hierarchies. In order to consider this construction as really
meaningful, it is necessary to study its stability under corrections
from higher dimensional terms on the branes as well as radiative
corrections. Another motivation for looking at deviations from
tri-bimaximal mixings is the possibility to get a non-zero
$\theta_{13}$, in case it turns out to be non-vanishing
experimentally. 
%Since there may be new sources of CP violation in the
%heavy neutrino sector $\theta_{13}\neq 0$ is not necessarily mandatory
%for a successful model of leptogenesis.

We focus first on the UV~brane and show that the effects of higher
dimensional A$_4$ invariants lead to the same pattern for the Majorana
mass matrix, except that some entries become complex.  This is the
only source of non-zero $\theta_{13}$ in this model. We start with
writing down the most general higher order operators allowed on the
UV~boundary: %%
\begin{equation}\label{eqn:UVcorrection}
-\delta\mathcal{L}_{UV}= \sum_{n\geqslant
2}\lambda_{n}\frac{\phi^n}{\Lambda^{n}}\psi_{\nu}\psi_{\nu}
+{\rm h.c.}
\end{equation}
with $n$ insertions of $\phi$. Now due to the Z$_2$-preserving
VEV $\langle\phi\rangle=(v,0,0)$, it is straightforward to show that
$\phi^3$ transforms as $\phi$ under A$_4$. Thus all the higher order
effects that cannot be reabsorbed into a redefinition of the lowest
order parameters arise from one operator: %%
\begin{equation}
-\delta\mathcal{L}_{UV}=\lambda_{2}\frac{\phi^2}{\Lambda^2}\psi_{\nu}\psi_{\nu}
+{\rm h.c.}
\end{equation}
This term contains actually three independent A$_4$ invariants which
lead to complex diagonal elements of the Majorana mass matrix and
the pattern of (\ref{eqn:MR}) is to be replaced by: %%
\begin{equation}
\mathbf{\mathcal{M}_M^\nu} =  F^2_{-\nu} R^{-1}
\left(\begin{matrix}
\epsilon_s+\delta_1 & 0 & 0 \\
0 & \epsilon_s+\delta_2 & \epsilon_t\\
0 & \epsilon_t & \epsilon_s+\delta_2^*
\end{matrix}\right).
\end{equation}
with $\delta_i\sim\mathcal{O}(v^2/\Lambda^2)$ and $\delta_1$ real.
The complex entries induce both a deviation of $\theta_{12}$ from its
maximal value and a non-zero $\sin(\theta_{13})$ of
$\mathcal{O}(\delta)$.

On the IR~brane, we show that the higher order corrections can only
affect $\theta_{12}$, as long as Z$_3$ remains unbroken on this
boundary. Again we start with the most general higher dimensional
invariants which take the following form: %%
\begin{equation}
-\delta\mathcal{L}_{IR}= \sum_{i=e,\mu,\tau} \sum_{n\geqslant
2}\lambda'_{i,n}\bar{\Psi}_L\frac{\phi'^n}{\Lambda'^{n+1}} H
\Psi_i+\sum_{n\geqslant 1} \kappa_n \bar{\Psi}_L
\frac{\phi'^n}{\Lambda'^{n+1}} H\Psi_\nu+{\rm h.c.}
\end{equation}
Since $\langle\phi'\rangle=(v',v',v')$ is Z$_3$ symmetric,
$\phi'^2$ transforms as $1+\phi'$. Then non-trivial
corrections reduce to terms with only one $\phi'$ insertion: %%
\begin{equation}
\label{eqn:IRcorrection}
-\delta\mathcal{L}_{IR}
=\frac{\kappa_1}{\Lambda'^2}\bar{\Psi}_L \phi' H \Psi_\nu
+{\rm h.c.}
\end{equation}
Since it is suppressed by only one power of $(v'\Lambda')$ compared to
the lowest order terms, this operator may easily destabilize the HPS
pattern. The extra Z$_2$ flavor symmetry is useful here, as we now
discuss. We choose the A$_4$ triplets odd under this additional Z$_2$,
and then this operator is forbidden by the Z$_2$ symmetry. However,
since $\phi'^2=1+\phi'$, the next Z$_2$ even operator generates the
same type of correction as the one linear in $\phi'$ but with a higher
suppression factor. In that case (\ref{eqn:IRcorrection}) has to be
replaced by
\begin{equation}
-\delta\mathcal{L}_{IR}=\frac{\kappa_2}{\Lambda'^3}\bar{\Psi}_L \phi'^2 H \Psi_\nu.
\end{equation}
This operator corrects the Dirac neutrino mass matrix by
introducing off diagonal elements of $\mathcal{O}(
v'^2/\Lambda'^2)$. As this term contains three independent
A$_4\times$Z$_2$ invariants, the diagonal Dirac matrix of
(\ref{eqn:mDnu}) becomes: %%
\begin{equation}
\mathbf{\mathcal{M}_{D}^\nu} =   y_\nu f_L f_{-\nu}
\frac{v_H}{\sqrt{2} R' \Lambda'}
\left(\begin{matrix}
1+\epsilon_1&\epsilon_2&\epsilon_3\\
\epsilon_3&1+\epsilon_1&\epsilon_2\\
\epsilon_2&\epsilon_3&1+\epsilon_1
\end{matrix}\right),
\end{equation}
with $\epsilon_i\sim\mathcal{O}(v'^2/\Lambda'^2)$. One can easily
check (see Appendix~B) that only $\sin(\theta_{12})$ is affected and
its deviation from the HPS prediction is of
$\mathcal{O}(v'^2/\Lambda'^2)$.

There is another important feature we want to stress at that point,
which is the fact that there is no correction to the charged lepton
mass matrix. This means that the $\Psi_{e,\mu,\tau}$ fields need not
be rotated again to get to the diagonal charged leptons basis even in
the presence of these higher order terms. As we shall see in section
5, an immediate consequence is the absence of tree level LVF in this
model, even when higher dimensional operators on are considered.

\begin{figure}[t]
\begin{center}
\begin{tabular}{ccccc}
\includegraphics[scale=0.5]{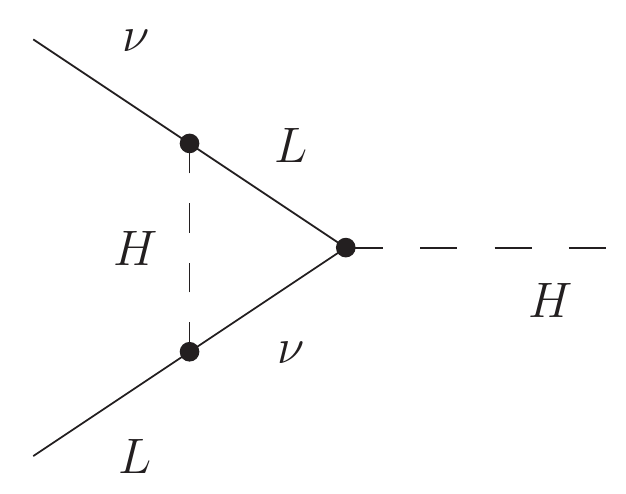}& &\includegraphics[scale=0.5]{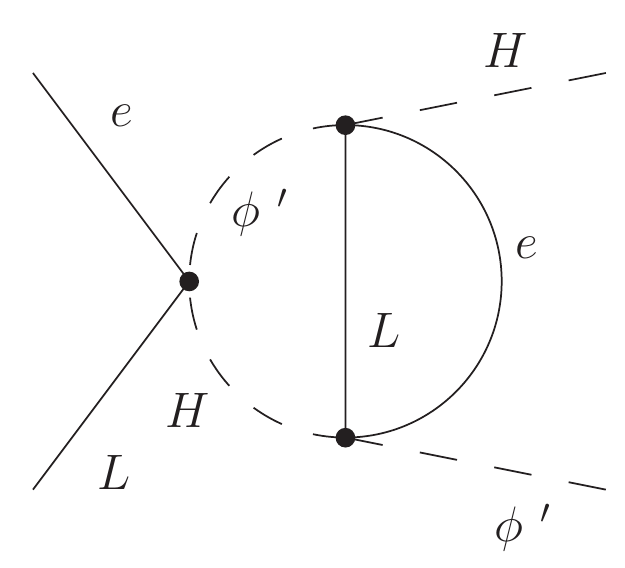}& & \includegraphics[scale=0.5]{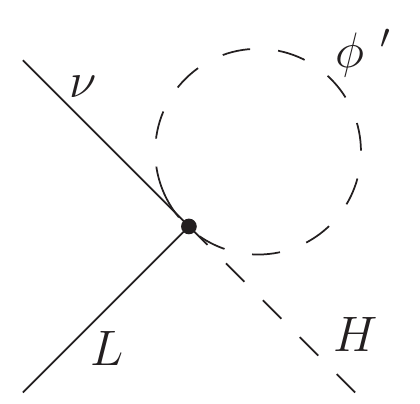}\\
&&&&\\
(a) & & (b) & & (c)
\end{tabular}
\end{center}
\caption{\label{fig:NDAdiagIR} Loop diagrams correcting the operators localized on the IR~brane which are relevant for NDA estimates.}
\end{figure}

We close this section by presenting an NDA estimates for the
allowed sizes of the IR~brane localized operators. These bounds are
important for estimating how large the deviations from the
HPS mixing matrix could actually be. For this purpose, we write down
again the most general Lagrangian on the IR~brane
\begin{equation}
-{\cal L}_{IR}=
\frac{y_\nu}{\Lambda'} \bar{\Psi}_L H \Psi_\nu
+\frac{y_{e,\mu,\tau}}{\Lambda'^2} \bar{\Psi}_L H \phi' \Psi_{e,\mu,\tau}
+\frac{\kappa_2}{2\Lambda'^3} \bar{\Psi}_L H {\phi'}^2\Psi_\nu
+{\rm h.c.}
\end{equation}
We require that the theory remains perturbative up to scale $E_N =
N m_{KK}$, which corresponds to the first $N^{\textrm{th}}$~KK modes
of the theory being weakly coupled. We should require $N\geq 3$, such
that at least the first few KK modes can be treated in perturbation
theory. We can then systematically require that by the time this
energy $E_N$ is reached all loop corrections are still smaller than
the lowest order terms. For example, for the first operator above
there is a one loop correction to the Yukawa vertex itself
(Fig. \ref{fig:NDAdiagIR}a), which implies
\begin{equation}
\frac{y_\nu^3}{16\pi^2} \frac{E_N^3}{\Lambda'^3} \leq
y_\nu\frac{E_N}{\Lambda'},
\end{equation}
where the linear running of the coupling has been taken into
account. Recalling $m_{KK}R'\simeq 2$, this implies for
$\Lambda'=R'^{-1}$ the perturbativity constraint
\begin{equation}
y_\nu \leq 2.
\end{equation}
Similarly, for the second operator there is a two-loop correction to
the tree-level operator (Fig. \ref{fig:NDAdiagIR}b), we get
\begin{equation}
\frac{y_e^3}{(16\pi^2)^2} \frac{E_N^6}{\Lambda'^6} \leq
y_e\frac{E_N^2}{\Lambda'^2} \ \ \to \ \ y_e \leq 4
\end{equation}
again for $\Lambda'=R'^{-1}$. Finally, the third operator gives
a one-loop correction to the first operator
(Fig. \ref{fig:NDAdiagIR}c), which implies the relation
\begin{equation}
\frac{\kappa_2}{16\pi^2}\frac{E_N^3}{\Lambda'^3} \leq y_\nu
\frac{E_N}{\Lambda'} \ \ \to \ \ \kappa_2\leq 4 y_\nu\leq 8.
\end{equation}
We can then estimate that the higher dimensional terms corrections to
$\sin(\theta_{12})$ are suppressed compared to the lowest order term
by at least $4 (v'R')^2$.

\begin{figure}[t]
\begin{center}
\begin{tabular}[ht]{ccccc}
\includegraphics[scale=0.5]{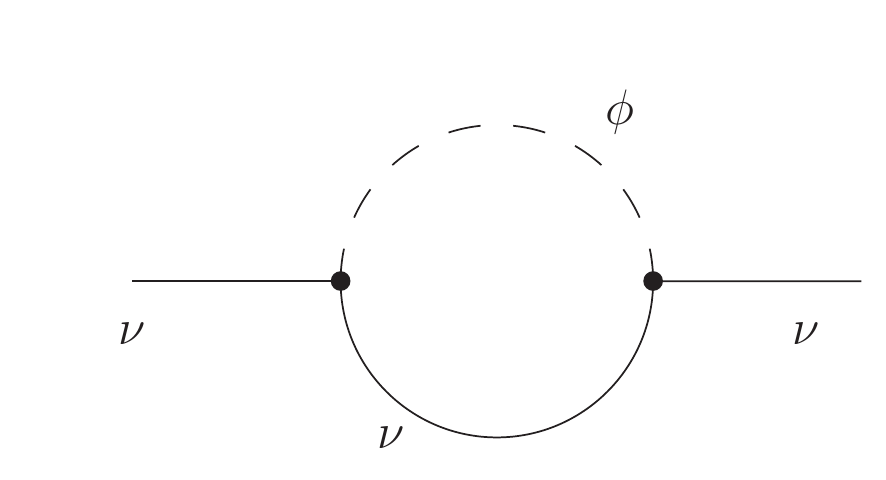}& &\includegraphics[scale=0.5]{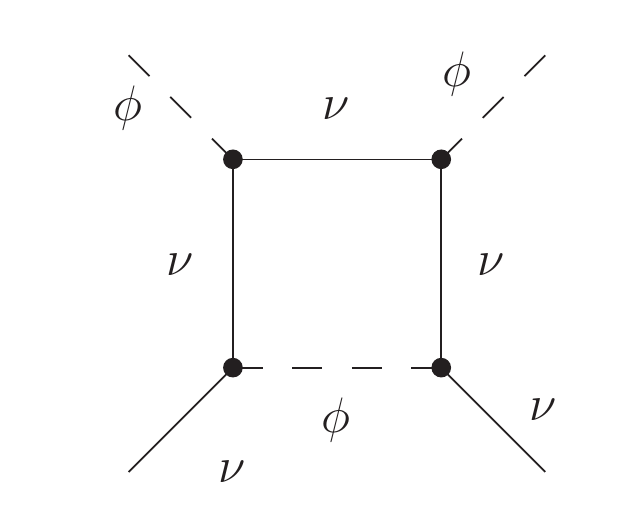}& & \includegraphics[scale=0.5]{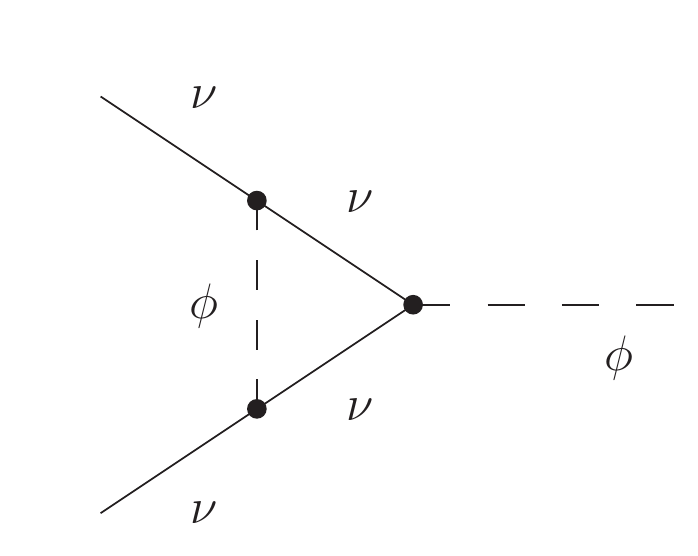}\\
&&&&\\
(a) & & (b) & & (c)\\
&&\includegraphics[scale=0.5]{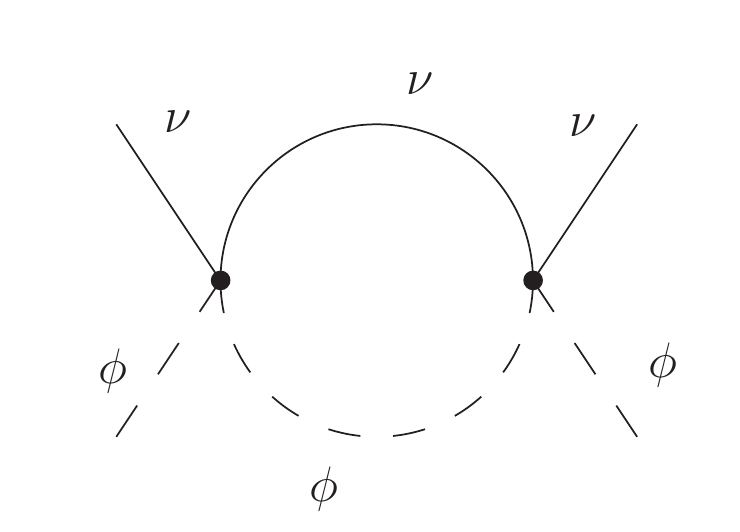}&&\\
&&&&\\
 & & (d) &&
\end{tabular}
\end{center}
\caption{\label{fig:NDAdiagUV} Loop diagrams illustrating the NDA estimates on the UV~brane.}
\end{figure}

We now apply the same arguments on the UV~brane where we have %%
\begin{equation}
-\mathcal{L}_{UV}=\frac{M}{2\Lambda}\psi_\nu\psi_\nu+x_\nu\frac{\phi}{2\Lambda}\psi_\nu\psi_\nu+\lambda_2\frac{\phi^2}{4\Lambda^2}\psi_\nu\psi_\nu
+\textrm{h.c.}
\end{equation}
We require that the theory remains perturbative up to the natural
scale on that boundary, namely until $E=R^{-1}$. First of all we focus
on the last operator which contribute at one loop to its own vertex
(Fig. \ref{fig:NDAdiagUV}d) which implies the usual constraint
\begin{equation}
\lambda_2\leq 4\pi,
\end{equation}
where we assumed $\Lambda=R^{-1}$. On the other hand, the second
operator induces one-loop corrections to the mass $M$, the third
operator, as well as its own vertex. From the diagrams of
Figs. \ref{fig:NDAdiagUV}a, \ref{fig:NDAdiagUV}b and
\ref{fig:NDAdiagUV}c we derive the following relations:
\begin{eqnarray}
x_\nu\leq 4\pi\sqrt{\epsilon_s},\qquad
x_\nu\leq
\lambda_2^{1/4}\sqrt{4\pi}\leq (4\pi)^{3/4}\sim 7,\qquad
 x_\nu\leq 4\pi.
\end{eqnarray}
We showed in the previous section that in order to reproduce the
observed hierarchical neutrino mass splittings, one must have
$\epsilon_s\sim\epsilon_t$ with $\epsilon_t=x_\nu(vR)$. Hence assuming
a not so small suppression factor $vR\sim 0.1$, such that
$\theta_{13}$ is not dramatically tiny, the first of the above
relations rewrites as $x_\nu\leq 16\pi^2 (vR)\sim 15$. Thus we end up
with the perturbative constraint
\begin{equation}
x_\nu\leq 7.
\end{equation}

%%%%%%%%%%%%%%%%%%%%%%%%%%%%%%%%%%%%%%%%%%%%%%%%%%%%%%
%%%%%%%%%%%%%%%%%%%%%%%%%%%%%%%%%%%%%%%%%%%%%%%%%%%%%%
\section{Numerical Scans and Electroweak Precision Bounds}
\label{sec:EW} \setcounter{equation}{0} \setcounter{footnote}{0}
%%%%%%%%%%%%%%%%%%%%%%%%%%%%%%%%%%%%%%%%%%%%%%%%%%%%%%
%%%%%%%%%%%%%%%%%%%%%%%%%%%%%%%%%%%%%%%%%%%%%%%%%%%%%%

One of the main constraints in models with new physics at the TeV
scale are the electroweak precision measurements (EWPM). For RS
models with custodial symmetry the generic bound on the KK scale is
about $m_{KK}\geq 3$ TeV, mostly from the contribution to the
S-parameter~\cite{ADMS}. Here we will check that the lepton sector
of our model indeed passes these tests for KK scales of order 3
TeV.~\footnote{It is possible to cancel the S-parameter by tuning
all LH bulk fermion masses to be $\sim 0.5$ \cite{fermiondeloc}, as
it is necessary in higgsless models~\cite{higgsless}. In this case
however one does not get any mass hierarchies and those need to be
introduced by hand as in~\cite{XDGim}.}

\begin{figure}[h!]
    \begin{center}
        \begin{tabular}{cc}
    \includegraphics[scale=0.2]{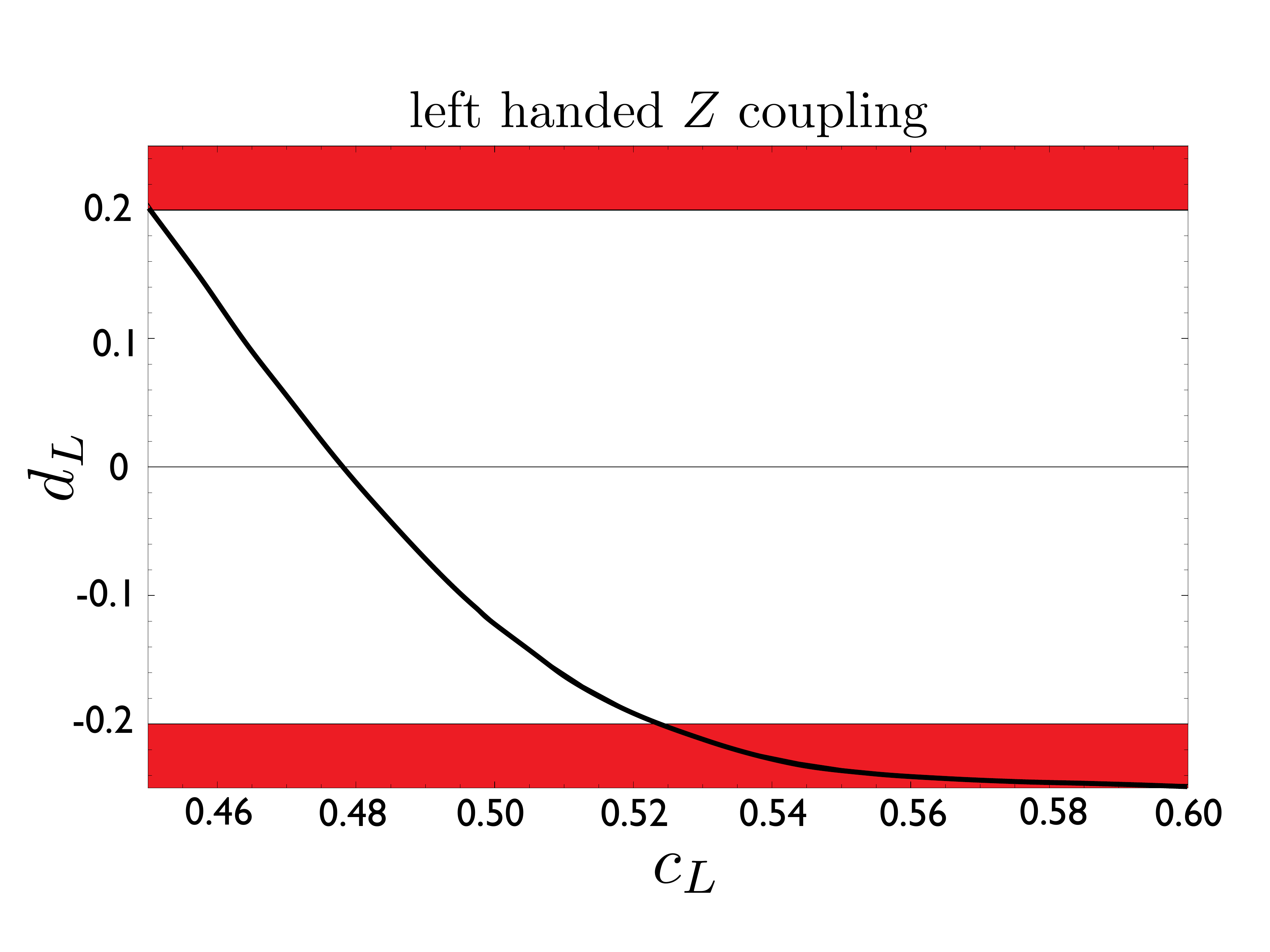}&\includegraphics[scale=0.2]{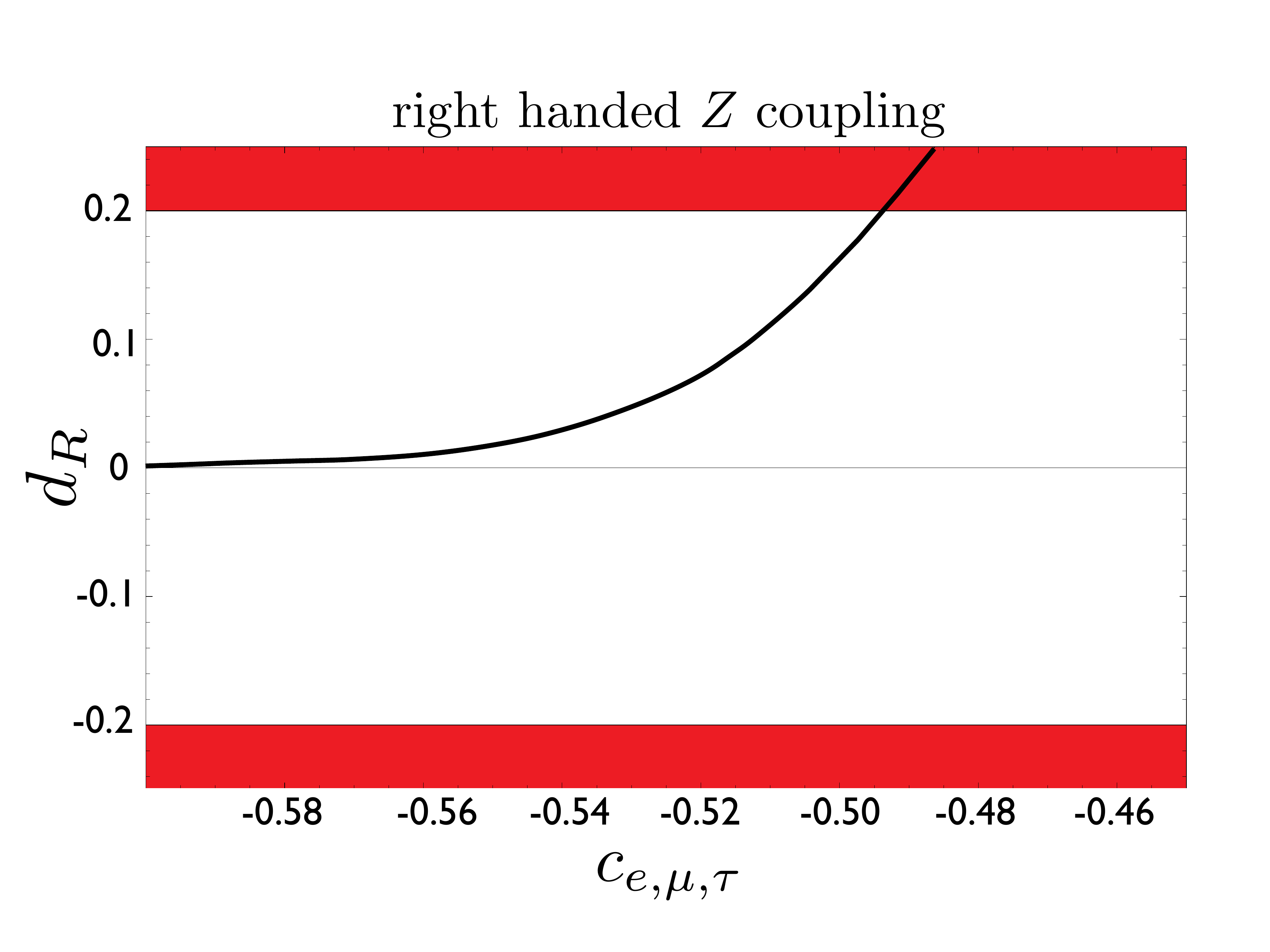}
        \end{tabular}
    \end{center}
    \caption{\label{fig:Zcpls}Deviation from SM Z couplings of $l_L$ and $l_R$ as function of $c$'s. $d_L=(g_Z^L-g_{Z,SM}^L)/g_{Z,SM}^L$ and $d_R=(g_Z^R-g_{Z,SM}^R)/g_{Z,SM}^R$ are plotted in per mil units. We took $R'^{-1}=1.5$ TeV and used $m_W=80.403$ GeV, $m_Z=91.1876$ GeV and $e=e(\mu=m_Z)=\sqrt{4\pi/128}$ as physical input observables. The red regions are excluded by EWPM in the leptonic sector.}
\end{figure}

The simplest way of checking the electroweak precision constraints
in an RS model with bulk fields is to canonically normalize the SM
gauge fields, and to determine the parameters $g_5$, $g_5'$ and
$v_H$ by requiring that the measured values of $M_W,M_Z$ and the
electromagnetic coupling $e$ are reproduced. This choice is somewhat
unconventional, since $M_W$ is less accurately measured than $G_F$,
however it simplifies the matching of the 5D parameters to the
observables significantly. In this scheme all corrections to
electroweak precision observables will be contained in the
fermion-gauge boson vertices, which can be simply calculated by wave
function overlaps, and compared to the SM predictions in terms of
the above input parameters, which for the lepton-$Z$-couplings are
\begin{eqnarray}
g^L_{Z,SM}&=&e\left(\frac{1}{2}-\frac{m_W^2}{m_Z^2}\right)\frac{m_Z}{m_W\sqrt{1-\frac{m_W^2}{m_Z^2}}},\\
g^R_{Z,SM}&=&e\frac{m_Z}{m_W}\sqrt{1-\frac{m_W^2}{m_Z^2}}.
\end{eqnarray}
 In a warped extra dimension the couplings of the left and right handed charged leptons are:
\begin{eqnarray}
g_Z^L&\simeq&\frac{1}{2}\int_R^{R'} dz \left(\frac{R}{z}\right)^4 \Big[\left[g_{5L} a^{L,3}(z,m_Z)+\tilde{g}_5 a^X(z,m_Z)\right] \chi_{c_L}(z)^2\Big]\\
g_Z^{R,i} &\simeq& \frac{1}{2}\int_R^{R'} dz \left(\frac{R}{z}\right)^4 \Big[\left[g_{5R} a^{R,3}(z,m_Z)+\tilde{g}_5 a^X(z,m_Z)\right] \psi_{c_i}(z)^2\Big].\label{eqn:Zleptons}
\end{eqnarray}
with $a(z,m_Z)$ the wave functions of the three neutral gauge fields
containing the physical $Z$, while the various $g_5$'s denote the
bulk gauge couplings\footnote{For completeness we review in appendix
\ref{sec:gaugebreak} how this quantities are calculated in terms of
physical observables.}. The origin of the deviations from the SM
predictions is the non-flatness of the $Z$ wave function. If the $Z$
boson was massless, its wave function would be flat and the lepton
couplings would become universal thanks to the orthonormality of
their wave functions. However as soon as the A$_4$ symmetry is
imposed, the left handed lepton couplings remain flavor blind. In
the ZMA the $Z$ coupling depends only on the bulk mass parameter
$c_L$. As usual we assume that the SM leptons are localized on the
UV brane, $c_L>1/2, c_{e,\mu ,\tau}<-1/2$. Since the $\tau$ is the
heaviest lepton, it has to be localized closest to the IR brane, so
it will be most sensitive to the non-flatness of the $Z$ close to
the IR brane, and so one expect the biggest deviations in the $\tau$
couplings.

The couplings of the charged leptons have been measured very
precisely at the LEP experiment~\cite{pdg}. Here we will require
that all lepton couplings are within 0.2\% of the SM prediction. The
plots of figure \ref{fig:Zcpls} show the deviation of the charged
lepton couplings to the $Z$ from their SM values as a function of
the $c$'s. We see that, while the $c_{e,\mu,\tau}$ can be as close
to $-1/2$ as required to reproduce the charged fermion hierarchy,
$c_L$ cannot depart too much from $1/2$ to remain within the
experimental bound. The fact that $c_L$ is preferred to be close to
$1/2$ may be surprising at first, but we remind the reader that
these vertex corrections also include the S-parameter. This is
actually a welcome fact, since in order to keep the $\tau$ Yukawa
coupling perturbative we have to take $c_L$ close to 1/2 anyway.
Thus we conclude from Fig.~\ref{fig:Zcpls} that with a 3 TeV KK mass
scale the electroweak precision constraints in the lepton sector are
safely satisfied.

\begin{figure}[h!]\label{fig:scan}
\centerline{\includegraphics[scale=0.29]{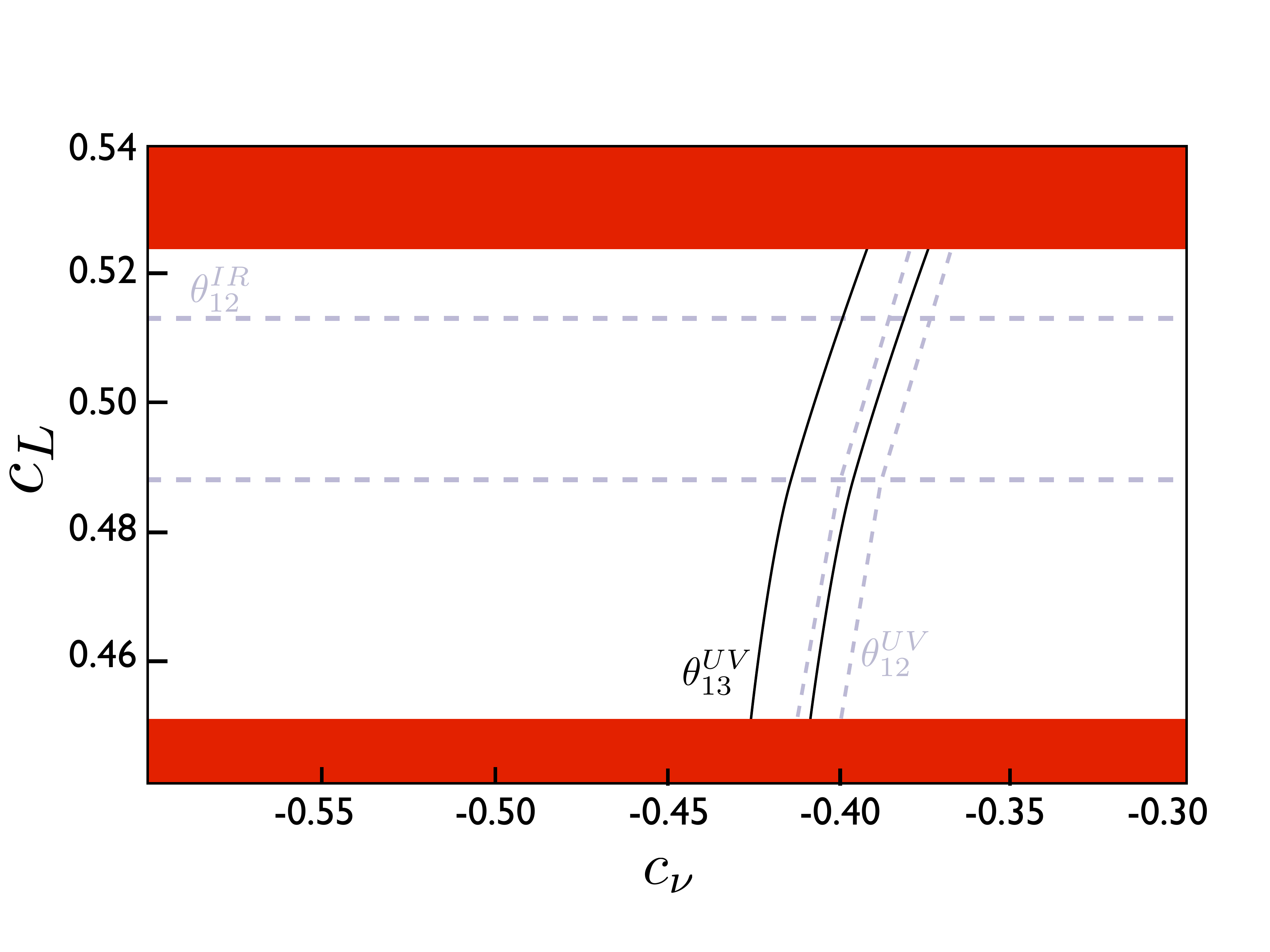}} \caption{Scan of
the parameter space reduced to $(c_L,c_\nu)$ as motivated in the
text. The red regions are excluded by electroweak precision
constraint on the $Z$ coupling. We then show within this region some
contours of the mixing angles delineating the largest values we
typically obtain in the presence of higher dimensional corrections.
The two horizontal contours are for the following values of
$\sin^2(2\theta_{12})={0.90,0.95}$ (from bottom to top) where only
the IR~higher dimensional operator is added with
$\epsilon_1=\epsilon_2=0$ and $\epsilon_3=8(v'R')^2$. The oblique
lines are contours of $\sin^2(2\theta_{13})={0.01,0.19}$ (from left
to right) and contours of $sin^2(2\theta_{12})={0.90,0.95}$ (from
left to right) generated by the higher order Majorana mass on the
UV~brane, for $\delta_1=\delta_2=0$ and $\delta_3=4\pi(vR)^2$.}
\end{figure}

Next we want to scan over the model's parameters and show explicitly
that the neutrino mixing angles, which deviate from the HPS pattern
in the presence of higher order operators, are within the allowed
range of the existing results of the neutrino experiments. Once the
usual RS solution to the hierarchy problem is imposed, we still have
12 free parameters in our setup: 5 $c$'s, 5 Yukawas and 2 VEVs
relevant for lepton masses $v,v'$, while adding higher order
operators brings 6 more coupling constants. We use the measured
lepton masses and the best fit neutrino mass splittings to fix 5 of
the lowest order parameters, which leaves still a lot of freedom to
explore. Therefore we add some mild assumptions in order to simplify
the parameter space and to try to only focus on the main predictions
without having to go into the details of the structure of the higher
dimensional operators. First of all we impose $c_\tau=-c_L$ and keep
$c_L$ as a free parameter. We also take the Yukawas on the IR brane
to saturate the perturbative bounds: $y_{e,\mu,\tau}=2y_\nu=4$. Thus
imposing $m_\tau=1.77$ GeV, $m_e=0.511$ MeV and $m_\mu=106$ MeV in
turn fixes $v'$ and $c_{e,\mu}$ as functions of $c_L$. Hence all IR
brane effects will be encoded in one parameter $c_L$, which is
constrained by the EWPM as discussed above. On the UV brane the
solar and atmospheric splittings fix the ratio $M/v$ and the overall
neutrino mass scale. Then taking $x_\nu=1$ we end up with
$v=v(c_\nu)$ as the only free parameter on this boundary.
Furthermore, $(vR)^2$ controls the size of $\theta_{13}$ generated
through higher order corrections. Finally we present in figure
\ref{fig:scan} some contours of the mixing angles in the plane
$(c_L,c_\nu)$ when one among the possible combinations of higher
order invariants are included on both branes. We considered
separately the effect on $\theta_{12}$ from the IR brane and the
predictions for all angles from the UV brane. In order to
demonstrate the robustness of the tri-bimaximal pattern under higher
order corrections in this model we have selected the worst case
where these operators saturate their perturbative limits, namely for
$\lambda_2=4\pi$ and $\kappa_2=8$. We conclude from Fig.
\ref{fig:scan} that, even when the deviations from HPS angles are
the largest possible, there is still a viable region satisfying the
constraints $\sin^2(2\theta_{13})<0.19$ (90CL),
$\sin^2(2\theta_{23})>0.92$ (90CL) and
$0.73\leq\sin^2(2\theta_{12})\leq 0.95$
($3\sigma$)~\cite{pdg,Maltoni}. Obviously the smaller the higher
order terms are the closer one would get to the HPS pattern. 
%%

%%%%%%%%%%%%%%%%%%%%%%%%%%%%%%%%%%%%%%%%%%%%%%%%%%%%%%
%%%%%%%%%%%%%%%%%%%%%%%%%%%%%%%%%%%%%%%%%%%%%%%%%%%%%%
\section{Constraints from Lepton Flavor Violation}
\label{sec:LFV} \setcounter{equation}{0} \setcounter{footnote}{0}
%%%%%%%%%%%%%%%%%%%%%%%%%%%%%%%%%%%%%%%%%%%%%%%%%%%%%%
%%%%%%%%%%%%%%%%%%%%%%%%%%%%%%%%%%%%%%%%%%%%%%%%%%%%%%

Flavor models usually predict new sources of flavor violations, and
so are only viable when the flavor scale is pushed to very high
values. The flavor scale for quarks in the usual anarchic RS flavor
models is around 20 TeV~\cite{CFW}. Thus it is very important to
understand what the possible sources of lepton flavor violation
(LFV) are in this model of lepton masses. Generically there are two
types of LFV sources: tree-level LFV via the exchange of heavy
neutral particles (like Z') or off-diagonal Z couplings, or loop
induced rare decays via charged current interactions.

Tree-level LFV operators have been considered in~\cite{AP}, and it
was found that the lepton flavor scale is at least 5 TeV, and larger
for smaller brane Yukawa couplings. We will first show here that the
structure of the A$_4$ symmetry used here is such that all
tree-level sources of lepton flavor violation are absent in this
theory. This can be seen in the following way. By the choice of the
A$_4$ representation the wave functions of the left handed SM
fermions are flavor universal (since they have the same $c_L$). So
the only source of flavor violations in the charged lepton sector is
the choice of different $c_{e,\mu ,\tau}$ necessary for the mass
hierarchies, and the brane Yukawa matrix (\ref{eq:mDe}). The
couplings to the neutral bulk fields like the KK tower of the Z are
controlled by the $c_{e,\mu ,\tau}$, and will be non-universal. Thus
any RH rotation to diagonalize the brane Yukawa matrix would induce
tree-level LFV's. However, we have seen that one of the magic
properties of the A$_4$ models is that the charged lepton mass
matrix is diagonalized by a single left handed rotation, and no
right handed rotation is necessary to diagonalize the mass matrix,
as described in (\ref{eq:mDediag}). The left-handed rotation is
harmless, since the bulk wave functions are universal in the LH
sector, while in the dangerous RH sector there is no rotation
necessary. This implies that there is a basis where the kinetic
terms and the mass terms for the charged leptons are simultaneously
diagonal, and so as a consequence there is no tree-level LFV in this
model.  The lepton embedding in A$_4$ provided us with the necessary
conditions to ensure the absence of LFV, namely universal left
handed $c$'s \textit{and} the absence of redefinition of the right
handed fields. Moreover we want to stress that this result remains
unchanged once higher dimensional brane operators are considered as
they were shown not to affect the rotation matrices of the charged
leptons.

 Thus all lepton flavor violation arises from charged
current interactions. In the SM loops involving neutrinos give
extremely small contributions to rare decays, however in the
presence of heavy KK neutrinos this is no longer the case. For
example in the case of the original neutrino mass model of~\cite{GN}
the large splittings of the heavy neutrinos, together with their
unsuppressed couplings to the SM fields yield a large loop-induced
$\mu\to e \gamma$ rate. In fact, Kitano~\cite{Kitano} showed that a
bound of $m_{KK}>25$ TeV applies in this case. In our case however
there is a generic softening of this bound, due to the fact that the
SM fermions are now localized close to the UV brane, and therefore
the charged current interactions with the KK neutrinos will be
suppressed. This will happen generically in any model with bulk
fermions. For the particular case of the A$_4$ model the situation
is even better: since the second Yukawa coupling of the charged
leptons involving the right handed neutrinos is uniform due to the
A$_4$ representations, the interactions with all  higgses (neutral
or charged) will be diagonalized in the same basis where the masses
are diagonalized. So one only needs to consider the exchange of
charged W's and their KK towers, together with KK neutrinos. In
fact, just as in the SM the loop induced contribution here will be
finite. The reason for that is that the allowed additional brane
localized counter term is of the form $L H \phi' \sigma^{\mu\nu} e_R
F_{\mu\nu}$, and this will be diagonalized once the charged lepton
mass matrix is diagonal. This is again a specific property of the
structure of the A$_4$ invariants.

Next we will give a rough estimate of the KK mass bound from these
processes. We will focus on loop induced $\mu\to e\gamma$ decays via
to exchange of a W-bosons and KK neutrinos. The branching fraction
from the exchange of a heavy neutrino and a W was calculated by
Cheng and Li~\cite{ChengLi} and is given by (assuming the coupling
to the W given by the usual SM gauge coupling $g$):
\begin{equation}
B(\mu\to e\gamma ) = \frac{3\alpha}{8\pi}
\left| \sum_i U_{\mu i}^* U_{ei}
F(\frac{m_i^2}{M_W^2}) \right|^2
\end{equation}
where the sum over i indicates the sum over a generation of KK
fermions, $U$ is the PMNS mixing matrix between the SM charged
leptons and a generation of KK neutrinos, and the function $F$ is
given by
\begin{equation}
F(z)= \frac{1}{6(1-z)^4} (10-43 z +78z^2-49 z^3+18 z^3\log z +4
z^4).
\end{equation}

First we specify to the case of the exchange of a SM W and a KK mode
neutrino. For $z\gg 1$ the approximate expression is $F(z) \approx
\frac{2}{3} +3 \frac{\log z}{z}$. Also, the coupling between the
zero mode SM fermions, a KK neutrino and the zero mode W is
suppressed at least by one factor of $f_L$, so there is an overall
$f_L^4$ appearing in the rate. This is the main difference compared
to the analysis of~\cite{Kitano}: there all SM fermions were
localized on the TeV brane, so the interaction with a KK neutrino
was unsuppressed. The leading term drops out due to the unitarity of
$U$, and so we are left with the approximation
\begin{equation}
B(\mu\to e \gamma ) < f_L^4 \frac{54 \alpha}{\pi}
\frac{m_W^4}{m_{KK}^4} \frac{\delta m_{KK}^2}{m_{KK}^2} \log^2
\frac{m_{KK}}{m_W},
\end{equation}
where $\delta m_{KK}$ is the characteristic splitting among the KK
modes in a given generation, given by $\frac{\delta m_{KK}}{m_{KK}}
\approx \frac{y^2 v_H^2}{2 m_{KK}^2}$. We find, that even for Yukawa
coupling close to the perturbative limit $y\sim 3$, and $c_L$ close
to the composite case $c_L=0.5$ the branching ratio is two orders of
magnitude below the experimental bound of $10^{-11}$ for a KK mass
scale of 3 TeV.

A slightly bigger contribution is obtained using the diagram where
in addition to the KK neutrino one exchanges a KK W. In this case
the gauge coupling could be as large as $g f_L \sqrt{\log R'/R}$.
One needs to take the function $F(z)$ at values $z \sim 1$ for which
it is approximately given by $F(z\sim 1) \sim \frac{17}{12}
+\frac{3}{20} (1-z)$. The universal piece drops out again due to the
unitarity of $U$ and we get an upper bound to the resulting
branching fraction of order
\begin{equation}
B(\mu\to e \gamma ) < f_L^4 \frac{27 \alpha}{800 \pi}
\frac{m_W^4}{m_{KK}^4} \frac{\delta m_{KK}^2}{m_{KK}^2} \log^2
\frac{R'}{R},
\end{equation}
which again numerically is smaller than $10^{-13}$ for a 3 TeV KK
mass.   Clearly since the numerical contributions turn out to be not
that far from experimentally interesting region it would be very
interesting to perform a more detailed calculation of the $\mu\to
e\gamma$ branching fraction in this model, including complete sums
over KK towers (and also in general RS models with bulk fermions and
Majorana neutrinos).

%%%%%%%%%%%%%%%%%%%%%%%%%%%%%%%%%%%%%%%%%%%%%%%%%%%%%%
%%%%%%%%%%%%%%%%%%%%%%%%%%%%%%%%%%%%%%%%%%%%%%%%%%%%%%
\section{Conclusions}
\label{sec:conclusions} \setcounter{equation}{0}
\setcounter{footnote}{0}
%%%%%%%%%%%%%%%%%%%%%%%%%%%%%%%%%%%%%%%%%%%%%%%%%%%%%%
%%%%%%%%%%%%%%%%%%%%%%%%%%%%%%%%%%%%%%%%%%%%%%%%%%%%%%

Warped extra dimensions provide a successful framework for flavor
models:  hierarchies in the masses and the
mixing angles are naturally generated. Since neutrinos do not show hierarchies in the mixing
angles, and only a mild hierarchy in the mass spectrum, one should
introduce additional symmetries in the lepton sector. In this paper
we have augmented the lepton sector of the RS model with the most
successful and economical global symmetry used for neutrino mass
models, the discrete non-abelian group A$_4$. With appropriate
assignments of the A$_4$ representations we can naturally achieve a
successful lepton mixing pattern, while the charged lepton mass
hierarchy is implemented as usual in RS models via wave function
overlap. The A$_4$ symmetry also eliminates all tree-level sources
of LFV in the neutral current sector, and so significantly lowers
the bound on the KK mass scale. LFV appears only through charged
current loops, and we estimated that the rate of $\mu\to e\gamma$ is
safely below the current experimental bound. So the most significant
bounds on this model come from the EWP measurements, and as usual
with KK mass scales of order 3 TeV these corrections will be also
safely within the experimental bounds.

%%%%%%%%%%%%%%%%%%%%%%%%%%%%%%%%%%%%%%%%%%%%%%%%%%%%%%
%%%%%%%%%%%%%%%%%%%%%%%%%%%%%%%%%%%%%%%%%%%%%%%%%%%%%%
\section*{Acknowledgments}
\label{sec:ackn} \setcounter{equation}{0} \setcounter{footnote}{0}
%%%%%%%%%%%%%%%%%%%%%%%%%%%%%%%%%%%%%%%%%%%%%%%%%%%%%%
%%%%%%%%%%%%%%%%%%%%%%%%%%%%%%%%%%%%%%%%%%%%%%%%%%%%%%

We thank Tony Gherghetta, Matt Reece and Andi Weiler for useful
discussions, and to Andi Weiler for comments on the manuscript. The
research of C.C. and Y.G. is supported in part by the NSF grant
PHY-0355005. C.C. was supported in part by the NSF grant PHY05-51164
at the KITP. The research of C.D. and C.G. is supported in part by
the RTN European Program MRTN-CT-2004-503369, by the EU FP6 Marie
Curie RTN ``UniverseNet" (MRTN-CT-2006-035863) and by the CNRS/USA
exchange grant 3503.

%%%%%%%%%%%
\begin{appendix}

%%%%%%%%%%%%%%%%%%%%%%%%%%%%
\section{Summary of A$_4$ group theory}\label{sec:A4}

$A_4$ is an $SO(3)$ subgroup which leaves the tetrahedron invariant.
It has 12 elements, two generators ($S$, $T$) connecting all of
them, and four irreducible representations: one three-dimensional
($3$) and three one-dimensional ($1$, $1'$ and $1''$, with
$(1')^*=1''$). Their products decompose as: %%
\begin{equation}
    \label{eqn:A4table}
\begin{array}{l}
1'\times  1''=1, \ \ \ 1 \times 3 = 3\\
1' \times 1' =1'', \  \ \ 1' \times 3 = 3\\
1'' \times 1'' =1', \  \ \ 1'' \times 3 = 3\\
3_x \times 3_y = 3_1+3_2+1+1'+1''
\end{array}\end{equation}
where for the last line, with $3_x=(x_1,x_2,x_3)$,
$3_y=(y_1,y_2,y_3)$ and working in a basis where the
three-dimensional representation of $S$ is diagonal: %%
\begin{eqnarray}
S=\left(\begin{matrix}
    1&0&0\\
    0&-1&0\\
    0&0&-1
 \end{matrix}\right),
 &&T=\left(\begin{matrix}
                    0&1&0\\
                    0&0&1\\
                    1&0&0
                     \end{matrix}\right),
\end{eqnarray}
one has: %%
\begin{eqnarray}
&& 1=x_1y_1+x_2y_2+x_3y_3\\
&& 1'=x_1y_1+\omega^2 x_2y_2+\omega x_3y_3\\
&& 1''=x_1y_1+\omega x_2y_2+\omega^2 x_3y_3\\
&& 3_1=(x_2y_3,x_3y_1,x_1y_2)\\
&& 3_2=(x_3y_2,x_1y_3,x_2y_1)
\end{eqnarray}
with $\omega=e^{2\pi i/3}$ the cubic root of unity, satisfying: %%
\begin{eqnarray}
1+\omega+\omega^2=0,&&\omega^*=\omega^2.
\end{eqnarray}
Note also that one has: $3\times 1'=3\sim u(x_1,\omega x_2,\omega^2
x_3)$, where $u\sim 1'$. The same holds for $3\times 1''$ with
$\omega\rightarrow \omega^2$.

%%%%%%%%%%%%%%%%%%%%%%%%%%%%%%
\section{$\theta_{13}=0$ and $\theta_{23}=\pi/4$ at any order on IR brane}

We explicitly show in this section that the higher dimensional operators on the IR brane only affect $\theta_{12}$. We recall that the Dirac mass matrix on the IR brane in presence of higher orders is of the form:
\begin{equation}
\mathbf{\mathcal{M}_{D}^\nu} =\left(\begin{matrix}
                         \alpha & \beta & \gamma\\
                         \gamma & \alpha & \beta\\
                         \beta & \gamma & \alpha
        \end{matrix}\right)
\end{equation}
while at lowest order the Majorana mass matrix on the UV is:
\begin{equation}
\mathbf{\mathcal{M}_M^\nu} = \left(\begin{matrix}
                    a & 0 & 0 \\
                    0 & a & d\\
                    0 & d & a
                                \end{matrix}\right).
\end{equation}
After the seesaw the Majorana mass matrix for the left-handed neutrinos (in the basis of diagonal charged leptons) shows the following pattern:
\begin{equation}
\mathbf{\tilde{\mathcal{M}}_\nu}=\left(\begin{matrix}
        b & c & c^*\\
        c & g & f\\
        c^* & f & g^*
                                       \end{matrix}\right).
\end{equation}
Note that even for real input parameters this matrix has complex entries due to $\omega=e^{2i\pi/3}$ factors that do not cancel out when the left-handed doublet is rotated with $\mathbf{V}$. If $c$ and $g$ were real, the Majorana mass matrix would be diagonalized with $\theta_{13}=0$ and $\theta_{23}=\pi/4$. In general it is a $3\times 3$ complex symmetric matrix which thus contains 12 independent parameters. They are the 3 real eigenvalues, the 3 mixing angles and 6 phases. Moreover one can always redefine the neutrino fields to absorb 3 of them, leaving only 2 Majorana phases and a CKM one. Thus with the redefinition $\nu_i\rightarrow e^{i\phi_i}\nu_i$ the mass matrix becomes:
\begin{equation}
\mathbf{\tilde{\mathcal{M}}_\nu}=\left(\begin{matrix}
        b\, e^{2i\phi_1} & |c| e^{i(\phi_c+\phi_1+\phi_2)} & |c| e^{i(\phi_1+\phi_2-\phi_c)}\\
        |c| e^{i(\phi_c+\phi_1+\phi_2)} & |g| e^{i(2\phi_2+\phi_g)} & f e^{i(\phi_2+\phi_3)}\\
        |c| e^{i(\phi_1+\phi_2-\phi_c)} & f e^{i(\phi_2+\phi_3)} & |g| e^{i(2\phi_3-\phi_g)}
                                       \end{matrix}\right).
\end{equation}
It is now not difficult to see that this matrix can be made real with $\phi_1=0$, $\phi_2=-\phi_3$ and $\phi_3=\phi_c$ provided that the relation $2\phi_c=\phi_g$ holds. Although the expressions of these phases in terms of the input parameters are quite cumbersome, we checked that the latter relation is satisfied in our model. Therefore we conclude that the most general higher dimensional corrections on the IR brane can only modify $\theta_{12}$ from its HPS value.\\
\end{appendix}

%%%%%%%%%%%%%%%%%%%%%%%%%%%%%%
\section{Review of Gauge Breaking via BCs}\label{sec:gaugebreak}
%%%%%%%%%%%%%%%%%%%%%%%%%%%%%%
We shortly review here how the bulk $SU(2)_L\times SU(2)_R\times U(1)_{B-L}$ gauge symmetry is broken in this setup. The main goal is to define how to get the gauge boson profiles as we will need them to compute the W and Z couplings to the standard model fermions. We note $A^L$, $A^R$ and $A^X$ as well as $g_{5L}$, $g_{5R}$ and $\tilde{g}_5$ the gauge fields and coupling constants associated with $SU(2)_L$, $SU(2)_R$ and $U(1)_{B-L}$ respectively. On the UV brane, the gauge symmetry breaks down to the SM gauge group $SU(2)_L\times U(1)Y$ from Dirichlet BC\footnote{This can be thought as being the result of a Higgs mechanism in the limit where the localized scalar is decoupled.} for the $SU(2)_R$ fields:
\begin{equation}
z=R : \left\{\begin{matrix}
        A^{R,\pm}_\mu=\partial_z A^{L,\pm}_\mu=0\\
        \partial_z A^{L,3}_\mu=0\\
        \tilde{g}_5 A^X_\mu-g_{5R}A^{R,3}_\mu=0\\
        \partial_z (g_{5R}A^X_\mu+\tilde{g}_{5}A^{R,3}_\mu)=0
             \end{matrix}\right.
\end{equation}
where $A^\pm\equiv(A^1\mp iA^2)/\sqrt{2}$. The chiral $SU(2)$'s are broken to the vectorial subgroup on the IR brane by the finite VEV of a Higgs bidoublet $\langle h\rangle=\mbox{diag}(v_H,v_H)/2$, and the resulting (mixed) BCs are:
\begin{equation}
z=R' : \left\{\begin{matrix}
        \partial_z(g_{5L}A^{L,a}_\mu-g_{5R}A^{R,a}_\mu)+\mathcal{V}(g_{5L}A^{L,a}_\mu-g_{5R}A^{R,a}_\mu)=0\\
        \partial_z (g_{5R}A^{L,a}_\mu+g_{5L}A^{R,a}_\mu)=0\\
        \partial_z A^X_\mu=0
             \end{matrix}\right.
\end{equation}
where $\mathcal{V}=(R'/R)(g_{5L}^2+g_{5R}^2)v_H^2/4$. The fifth components of the gauge fields have opposite BC. Obviously, these BCs allow for a massless (flat) mode which is nothing else but the photon field associated with the unbroken $U(1)$ of the (compactified) effective theory, and the KK decomposition is of the form:
\begin{eqnarray}\label{eqn:gaugeKKmodes}
A^{L,R,\pm}_\mu(x,z)&=&a^{L,R}(z,m_W)W^\pm_\mu(x)+\cdots\\
A^{L,R,3}_\mu(x,z)&=&\frac{\tilde{g}_5}{g_{5L,R}}a_0\gamma_\mu(x)+a^{L,R,3}(z,m_Z)Z_\mu(x)+\cdots\\
A^X_\mu(x,z)&=&a_0\gamma_(x)+a^X(z,m_Z)Z_\mu(x)+\cdots
\end{eqnarray}
where the $\cdots$ stand for heavier KK resonances, and the wave functions are given by $a(z,m)=z(A J_1(m z)+B Y_1(m z))$.\\
Yet remains the definition of the 5D gauge couplings. For this we have to match the latter on the 4D SM couplings. The fact that there is no $SU(2)_L\times U(1)_Y$ symmetry in the effective 4D action makes the definition of the SM couplings somehow arbitrary. As matching conditions, we choose to recover the measured values of $m_W$, $m_Z$ and the electric charge $e$. After having fixed the values of $R$, $R'$ and the ratio of the left/right gauge couplings, $r\equiv g_{5L}/g_{5R}$, three parameters remain to be determined by the matching procedure, namely $g_{5L}$, $\tilde{g}_5$ and $v_H$. To fit them we proceed as follows. First, we relate $\tilde{g}_5$ and $g_{5L}$ by imposing that the (canonically normalized) photon couples to the electric charge  $Q=T_{3,L}+T_{3,R}-Q_{B-L}/2$. Given the KK decomposition above we get:
\begin{equation}
\frac{1}{e^2}=\left(\frac{1+r^2}{g_{5L}^2}+\frac{1}{\tilde{g}_5^2}\right)R\log(R'/R).
\end{equation}
Then, the charged boson quantization equation fixes the product $g_{5L}v_H$ as a function of $m_W$. And plugging back these two relations into the neutral boson quantization equation allows to find the Higgs VEV as a function of $m_Z$, $m_W$ and $e$. Finally to fully determine the wave-functions we need to make the W and the Z are canonical fields by imposing:
\begin{eqnarray}
\int_R'^R dz \left(\frac{R}{z}\right)\left[a^{L,\pm}(z,m_W)^2+a^{R,\pm}(z,m_W)^2\right]&=&1\\
\int_R'^R dz \left(\frac{R}{z}\right)\left[a^{L,3}(z,m_Z)^2+a^{R,3}(z,m_Z)^2+a^X(z,m_Z)^2\right]&=&1.
\end{eqnarray}

%%%%%%%%%%%%%%%%%%%%%%%%%%%%%%%%%%%%%%%%%%%%%%%%%%%%%%%%%%%%%%%%%%%%


\begin{thebibliography}{9}

\bibitem{RS}
L.~Randall and R.~Sundrum,
  %``A large mass hierarchy from a small extra dimension,''
  Phys.\ Rev.\ Lett.\  {\bf 83}, 3370 (1999)
  [arXiv:hep-ph/9905221].
  %%CITATION = PRLTA,83,3370;%%




\bibitem{GN}
Y.~Grossman and M.~Neubert,
  %``Neutrino masses and mixings in non-factorizable geometry,''
  Phys.\ Lett.\  B {\bf 474}, 361 (2000)
  [arXiv:hep-ph/9912408].
  %%CITATION = PHLTA,B474,361;%%


\bibitem{GP}
T.~Gherghetta and A.~Pomarol,
  %``Bulk fields and supersymmetry in a slice of AdS,''
  Nucl.\ Phys.\  B {\bf 586}, 141 (2000)
  [arXiv:hep-ph/0003129].
  %%CITATION = NUPHA,B586,141;%%


\bibitem{AS}
 N.~Arkani-Hamed and M.~Schmaltz,
  %``Hierarchies without symmetries from extra dimensions,''
  Phys.\ Rev.\  D {\bf 61}, 033005 (2000)
  [arXiv:hep-ph/9903417].
  %%CITATION = PHRVA,D61,033005;%%

\bibitem{HS}
S.~J.~Huber and Q.~Shafi,
  %``Fermion masses, mixings and proton decay in a Randall-Sundrum model,''
  Phys.\ Lett.\  B {\bf 498}, 256 (2001)
  [arXiv:hep-ph/0010195];
  %%CITATION = PHLTA,B498,256;%%
S.~J.~Huber,
  %``Flavor violation and warped geometry,''
  Nucl.\ Phys.\  B {\bf 666}, 269 (2003)
  [arXiv:hep-ph/0303183].
  %%CITATION = NUPHA,B666,269;%%


\bibitem{APS}
K.~Agashe, G.~Perez and A.~Soni,
  %``B-factory signals for a warped extra dimension,''
  Phys.\ Rev.\ Lett.\  {\bf 93}, 201804 (2004)
  [arXiv:hep-ph/0406101];
  %%CITATION = PRLTA,93,201804;%%
%``Flavor structure of warped extra dimension models,''
  Phys.\ Rev.\  D {\bf 71}, 016002 (2005)
  [arXiv:hep-ph/0408134].
  %%CITATION = PHRVA,D71,016002;%%


\bibitem{pdg}
W.~M.~Yao {\it et al.}  [Particle Data Group],
  %``Review of particle physics,''
  J.\ Phys.\ G {\bf 33}, 1 (2006).
  %%CITATION = JPHGB,G33,1;%%










\bibitem{Gustavo}
G.~Burdman,
  %``Flavor violation in warped extra dimensions and CP asymmetries in B
  %decays,''
  Phys.\ Lett.\  B {\bf 590}, 86 (2004)
  [arXiv:hep-ph/0310144].
  %%CITATION = PHLTA,B590,86;%%






\bibitem{CFW}
 C.~Cs\'aki, A.~Falkowski and A.~Weiler,
  %``The Flavor of the Composite Pseudo-Goldstone Higgs,''
  arXiv:0804.1954 [hep-ph].
  %%CITATION = ARXIV:0804.1954;%%



\bibitem{HSleptons}
S.~J.~Huber and Q.~Shafi,
  %``Majorana neutrinos in a warped 5D standard model,''
  Phys.\ Lett.\  B {\bf 544}, 295 (2002)
  [arXiv:hep-ph/0205327];
  %%CITATION = PHLTA,B544,295;%%
%S.~J.~Huber and Q.~Shafi,
  %``Seesaw mechanism in warped geometry,''
  Phys.\ Lett.\  B {\bf 583}, 293 (2004)
  [arXiv:hep-ph/0309252].
  %%CITATION = PHLTA,B583,293;%%



\bibitem{AP}
K.~Agashe, A.~E.~Blechman and F.~Petriello,
  %``Probing the Randall-Sundrum geometric origin of flavor with lepton  flavor
  %violation,''
  Phys.\ Rev.\  D {\bf 74}, 053011 (2006)
  [arXiv:hep-ph/0606021].
  %%CITATION = PHRVA,D74,053011;%%

\bibitem{XDGim}
G.~Cacciapaglia, C.~Cs\'aki, J.~Galloway, G.~Marandella, J.~Terning
and A.~Weiler,
  %``A GIM Mechanism from Extra Dimensions,''
  JHEP {\bf 0804}, 006 (2008)
  [arXiv:0709.1714 [hep-ph]].
  %%CITATION = JHEPA,0804,006;%%

\bibitem{inprep}
C.~Cs\'aki, A.~Falkowski, A.~Weiler, to appear; C.~Cs\'aki,
Y.~Grossman, G.~Perez, Z.~Surujon, A.~Weiler, to appear.

\bibitem{MuChun}
M.~C.~Chen and H.~B.~Yu,
  %``Minimal Flavor Violation in the Lepton Sector of the Randall-Sundrum
  %Model,''
  arXiv:0804.2503 [hep-ph].
  %%CITATION = ARXIV:0804.2503;%%


\bibitem{FPR}
A.~L.~Fitzpatrick, G.~Perez and L.~Randall,
  %``Flavor from Minimal Flavor Violation & a Viable Randall-Sundrum Model,''
  arXiv:0710.1869 [hep-ph].
  %%CITATION = ARXIV:0710.1869;%%




\bibitem{Ma}
E.~Ma and G.~Rajasekaran,
  %``Softly broken A(4) symmetry for nearly degenerate neutrino masses,''
  Phys.\ Rev.\  D {\bf 64}, 113012 (2001)
  [arXiv:hep-ph/0106291];
  %%CITATION = PHRVA,D64,113012;%%
K.~S.~Babu, E.~Ma and J.~W.~F.~Valle,
  %``Underlying A(4) symmetry for the neutrino mass matrix and the quark  mixing
  %matrix,''
  Phys.\ Lett.\  B {\bf 552}, 207 (2003)
  [arXiv:hep-ph/0206292];
  %%CITATION = PHLTA,B552,207;%%
E.~Ma,
  %``A(4) origin of the neutrino mass matrix,''
  Phys.\ Rev.\  D {\bf 70}, 031901 (2004)
  [arXiv:hep-ph/0404199].
  %%CITATION = PHRVA,D70,031901;%%




%\cite{Altarelli:2005yp}
\bibitem{AF}
  G.~Altarelli and F.~Feruglio,
  %``Tri-bimaximal neutrino mixing from discrete symmetry in extra
  %dimensions,''
  Nucl.\ Phys.\  B {\bf 720} (2005) 64
  [arXiv:hep-ph/0504165];
  %%CITATION = NUPHA,B720,64;%%
 %``Tri-bimaximal neutrino mixing, A(4) and the modular symmetry,''
  Nucl.\ Phys.\  B {\bf 741}, 215 (2006)
  [arXiv:hep-ph/0512103].
  %%CITATION = NUPHA,B741,215;%%




\bibitem{HPS}
P.~F.~Harrison, D.~H.~Perkins and W.~G.~Scott,
  %``Tri-bimaximal mixing and the neutrino oscillation data,''
  Phys.\ Lett.\  B {\bf 530}, 167 (2002)
  [arXiv:hep-ph/0202074].
  %%CITATION = PHLTA,B530,167;%%



\bibitem{Kitano}
 R.~Kitano,
  %``Lepton flavor violation in the Randall-Sundrum model with bulk
  %neutrinos,''
  Phys.\ Lett.\  B {\bf 481}, 39 (2000)
  [arXiv:hep-ph/0002279].
  %%CITATION = PHLTA,B481,39;%%

\bibitem{ADMS}
K.~Agashe, A.~Delgado, M.~J.~May and R.~Sundrum,
  %``RS1, custodial isospin and precision tests,''
  JHEP {\bf 0308}, 050 (2003)
  [arXiv:hep-ph/0308036].
  %%CITATION = JHEPA,0308,050;%%



%\cite{Csaki:2003sh}
\bibitem{CGHST}
  C.~Cs\'aki, C.~Grojean, J.~Hubisz, Y.~Shirman and J.~Terning,
  %``Fermions on an interval: Quark and lepton masses without a Higgs,''
  Phys.\ Rev.\  D {\bf 70} (2004) 015012
  [arXiv:hep-ph/0310355].
  %%CITATION = PHRVA,D70,015012;%%

\bibitem{Cacciapaglia:2006mz}
  G.~Cacciapaglia, C.~Cs\'aki, G.~Marandella and J.~Terning,
  %``The gaugephobic Higgs,''
  JHEP {\bf 0702}, 036 (2007)
  [arXiv:hep-ph/0611358].
  %%CITATION = JHEPA,0702,036;%%
  
 \bibitem{fermiondeloc}
G.~Cacciapaglia, C.~Cs\'aki, C.~Grojean and J.~Terning,
  %``Curing the ills of Higgsless models: The S parameter and unitarity,''
  Phys.\ Rev.\  D {\bf 71}, 035015 (2005)
  [arXiv:hep-ph/0409126];
  %%CITATION = PHRVA,D71,035015;%%
R.~Foadi, S.~Gopalakrishna and C.~Schmidt,
  %``Effects of fermion localization in Higgsless theories and electroweak
  %constraints,''
  Phys.\ Lett.\  B {\bf 606}, 157 (2005)
  [arXiv:hep-ph/0409266].
  %%CITATION = PHLTA,B606,157;%%


\bibitem{higgsless}
 C.~Cs\'aki, C.~Grojean, L.~Pilo and J.~Terning,
  %``Towards a realistic model of Higgsless electroweak symmetry breaking,''
  Phys.\ Rev.\ Lett.\  {\bf 92}, 101802 (2004)
  [arXiv:hep-ph/0308038].
  %%CITATION = PRLTA,92,101802;%%



\bibitem{Maltoni}
  M.~Maltoni, T.~Schwetz, M.~A.~Tortola and J.~W.~F.~Valle,
  %``Status of global fits to neutrino oscillations,''
  New J.\ Phys.\  {\bf 6}, 122 (2004)
  [arXiv:hep-ph/0405172].
  %%CITATION = NJOPF,6,122;%%


\bibitem{ChengLi}
T.~P.~Cheng and L.~F.~Li,
  %``Nonconservation Of Separate Mu - Lepton And E - Lepton Numbers In Gauge
  %Theories With V+A Currents,''
  Phys.\ Rev.\ Lett.\  {\bf 38}, 381 (1977);
  %%CITATION = PRLTA,38,381;%%
 %T.~P.~Cheng and L.~F.~Li,
  %``Muon Number Nonconservation Effects In A Gauge Theory With V A Currents And
  %Heavy Neutral Leptons,''
  Phys.\ Rev.\  D {\bf 16}, 1425 (1977).
  %%CITATION = PHRVA,D16,1425;%%










\end{thebibliography}
\end{document}